\documentclass[useAMS,usenatbib]{mn2e}
\usepackage{epsf, epsfig}

\newcommand{\ltapprox}{\raisebox{-0.5ex}{$\,\stackrel{<}{\scriptstyle
\sim}\,$}}

\title[Discovery of two ULXs in NGC\,7424]{Multi-band study 
  of NGC\,7424 and its two newly-discovered ULXs}
\author[R. Soria et al.]{R. Soria$^{1,2}$\thanks{E-mail:
rsoria@cfa.harvard.edu}, 
Z. Kuncic$^{3}$, J. W. Broderick$^{3}$, and S. D. Ryder$^{4}$\\
$^{1}$Harvard-Smithsonian Center for Astrophysics, 
	60 Garden st, Cambridge, MA 02138, USA\\
$^{2}$Mullard Space Science Laboratory (UCL), Holmbury St Mary, 
	Dorking, Surrey, RH5 6NT, UK\\
$^{3}$School of Physics,
University of Sydney, NSW 2006, Australia\\
$^{4}$Anglo-Australian Observatory, PO Box 296, Epping, NSW 1710, 
Australia} 
\begin{document}

\date{Accepted 2006 May 24. Received 2006 May 22; 
in original form 2006 March 30.}


\maketitle


\begin{abstract}
We have studied the face-on, barred spiral NGC\,7424 
(site of the rare Type IIb SN\,2001ig) with 
{\it Chandra}, {\it Gemini} and the Australia Telescope 
Compact Array. After giving revised X-ray colours and 
luminosity of the supernova, here we focus on some other interesting 
sources in the galaxy: in particular, our serendipitous discovery 
of two ultraluminous X-ray sources (ULXs). The brighter one 
($\sim 10^{40}$ erg s$^{-1}$) has a power-law-like spectrum 
with photon index $\Gamma \approx 1.8$. The other ULX 
shows a spectral state transition or outburst between the 
two {\it Chandra} observations, 20 days apart. 
Optical data show that this ULX is located in 
a young (age $\approx 7$--$10$ Myr), bright complex rich with OB stars 
and clusters. An exceptionally bright, 
unresolved radio source ($0.14$ mJy at $4.79$ GHz,
implying a radio luminosity twice as high as Cas A)
is found slightly offset 
from the ULX ($\approx 80$ pc). Its radio spectral index 
$\alpha \approx -0.7$ suggests optically-thin synchrotron 
emission, either from a young supernova remnant or from a radio lobe 
powered by a ULX jet. An even brighter, unresolved radio source 
($0.22$ mJy at $4.79$ GHz) is found in another young, massive 
stellar complex, not associated with any X-ray sources: 
based on its flatter radio spectral index ($\alpha \approx -0.3$), 
we suggest that it is a young pulsar wind nebula, a factor of $10$ 
more radio luminous than the Crab.
\end{abstract}


\section{Introduction}

The physical interpretation of ultra-luminous 
X-ray sources (ULXs) remains a fervently debated 
and unsolved problem (for recent reviews, see King 2006; 
Fabbiano \& White 2006; Colbert \& Miller 2004).
They could be scaled-up equivalents of Galactic 
black hole (BH) X-ray binaries, with a higher-mass 
accretor (up to $\sim$ a few $10^2 M_{\odot}$ for 
the brightest sources, if the accretion is 
isotropic and Eddington-limited). 
Alternatively, they could be 
beamed sources---either mild geometrical beaming 
or relativistic Doppler boosting have been suggested  
(King et al.~2001; K\"{o}rding, Falcke \& Markoff~2002; Fabrika 
\& Mescheryakov 2001). A direct comparison with 
Galactic BH X-ray binaries is hampered by 
at least two problems. Firstly, the mass and 
evolutionary state of the donor star 
is unknown in almost all cases: some ULXs 
appear to be associated with OB associations 
or young star clusters; others, however, do not have 
optical counterparts down to typical detection 
limits $M_V \sim -5$ mag. 
Secondly, most ULXs have been observed in the X-rays 
only once or few times over the last decade; thus,
it is still unclear whether they follow the same  
patterns of spectral evolution and state transitions 
as daily-monitored Galactic BH X-ray binaries, 
and over what time-scale.

In this paper, we report the serendipitous discovery 
of two ULXs in the face-on SABcd galaxy NGC\,7424. 
Henceforth, we adopt a distance $d = 11.5$ Mpc (Tully 1988); 
a slightly lower distance of $10.9$ Mpc was 
estimated by B\"oker et al. (2002). The galaxy 
was observed twice with {\it Chandra}, on 
2002 May 21--22 and 2002 June 11, to study 
the X-ray supernova SN\,2001ig (Evans, White \& Bembrick 2001; 
Schlegel \& Ryder 2002). 
For the same reason, it was monitored 
on a regular basis for three years with 
the Australia Telescope Compact Array (ATCA) 
(Ryder et al.~2004). 
We studied the X-ray spectral state of the two ULXs 
found near SN\,2001ig, and their variability 
between the two observations. We then 
searched for their radio and optical counterparts, 
using the ATCA dataset, together with archival 
{\it HST}/WFPC2 and new {\it Gemini} images (Ryder, 
Murrowood \& Stathakis~2006)\footnote{High-resolution colour 
images of NGC\,7424 from the {\it Gemini} observations are 
available at http://www.gemini.edu/2001igpr}. Finally, 
we determined the X-ray luminosity and colours 
of the other bright X-ray sources detected in 
the field of this galaxy.

\section{X-ray data analysis}

We analysed the two {\it Chandra} datasets 
(Table 1) with standard tasks from 
the {\footnotesize CIAO}-3.3 data-reduction 
package\footnote{http://cxc.harvard.edu/ciao}.
Specifically, we checked for aspect offsets in the level-1 
event files, and applied the appropriate astrometric 
corrections. We created new bad-pixel files with 
the {\tt acis\_run\_hotpix} script, which flags hot 
pixels and cosmic-ray afterglows. We then created 
new level-2 event files with {\tt acis\_process\_events},  
which provides the newest gain maps, time-dependent 
gain corrections, charge-transfer-inefficiency corrections, 
pixel and PHA randomization. We then filtered the event files 
for bad grades and applied the good-time-intervals.
We checked that there were no significant background 
flares in either observations.

We used {\tt wavdetect} to identify discrete sources in both 
datasets, and in different energy bands. We also used  
a combined image (obtained with {\tt merge\_all}) 
to detect fainter sources. Lightcurves of the brightest 
sources were obtained with {\tt dmextract}.
Spectra and corresponding response files were 
extracted with {\tt specextract}, and modelled 
with {\footnotesize XSPEC} version 11.3.1 (Arnaud ~1996).

\begin{table}
\begin{tabular}{lccc}
\hline
Obs.~ID & Instrument & Start date & Exp. time (ks)\\
\hline
3495 & ACIS-S  & 2002-05-21 23:34:38 & 23.4\\
3496 & ACIS-S  & 2002-06-11 06:02:07 & 23.9\\
\hline
\end{tabular}
\caption{Details of the {\it Chandra} 
observations of NGC\,7424 used for our study.}
\end{table}

\begin{figure*}
\epsfig{figure=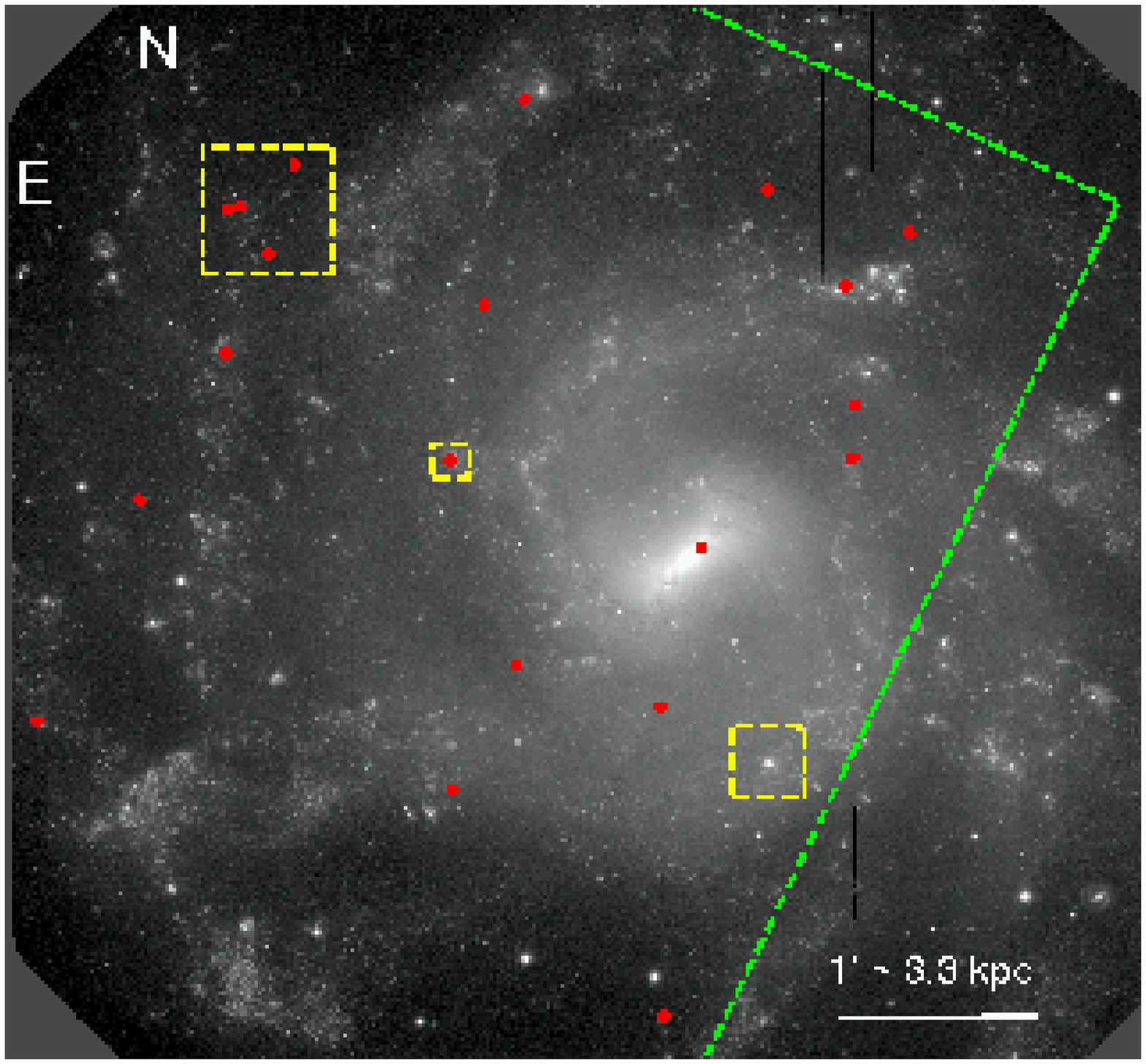, width=9.0cm}
\epsfig{figure=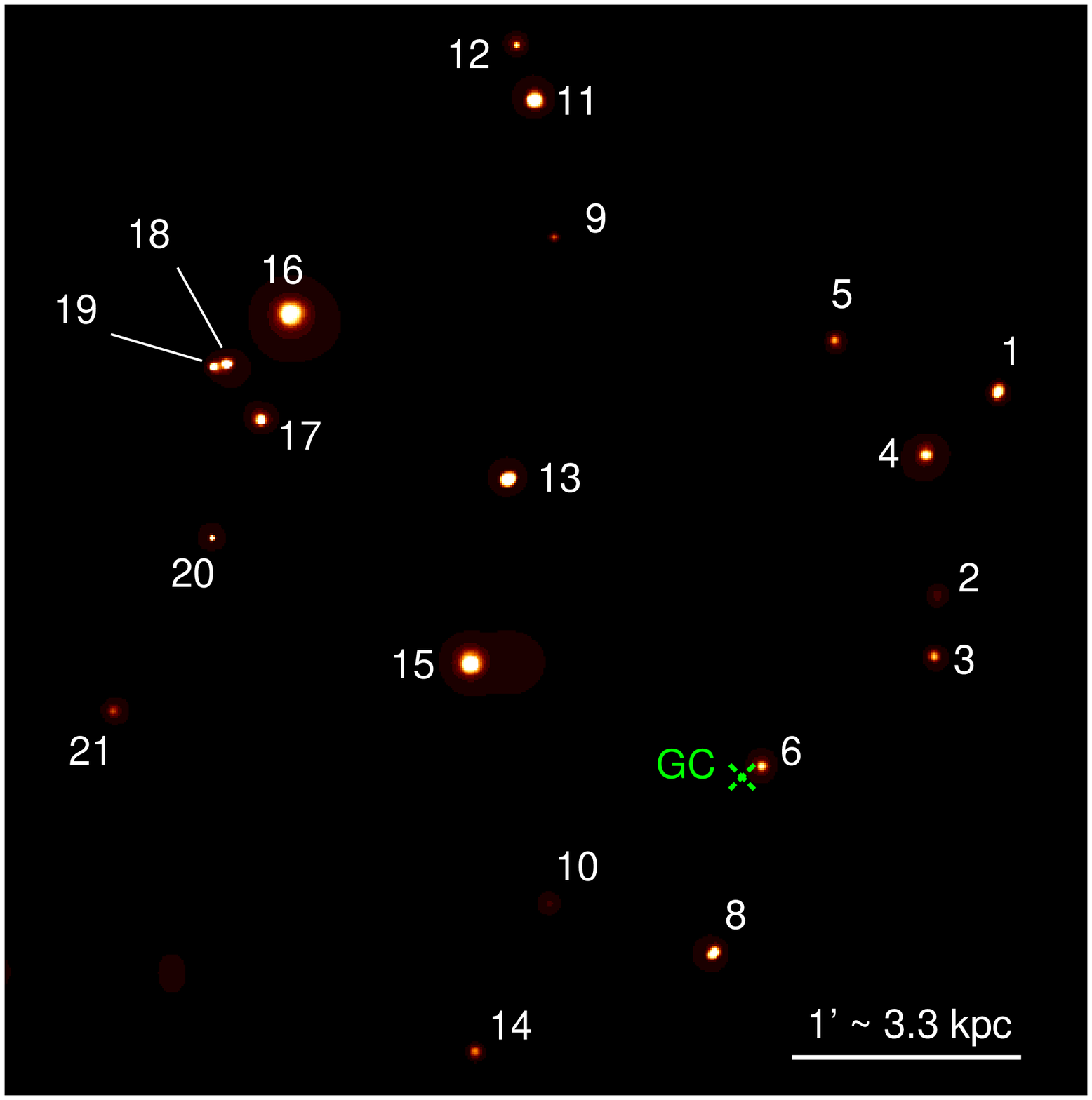, width=8.45cm}
\caption{Map of the X-ray sources detected by {\it Chandra}. 
Left: source location (red circles) plotted over a {\it Gemini} $r'$ image. 
The three small yellow insets identify regions that are 
discussed in the text and shown in more detail later 
(Figures 8, 10 and 11 respectively). The larger green wedge marks the boundary 
of the ACIS-S3 chip. Right: smoothed {\it Chandra} image 
in the $0.3$--$8$ keV band, with the source identification 
referring to Table 2. The galactic center is marked by a green 
cross, close to but not coincident with a supersoft source (No. 6).}
\end{figure*}

\section{X-ray study: main results}

\subsection{Overview}

From the combined 47.3 ks image, we identified 
22 point-like sources inside the D$_{25}$ ellipse 
(Table 2 and Figure 1), at a significance $> 3\sigma$ 
in the $0.3$--$10$ keV energy band. 
In fact, the ACIS field of view covers only 
$\sim 2/3$ of that region: this is partly because 
the S3 chip is centered on the location of SN\,2001ig 
rather than the galactic nucleus; however, 
the nucleus is also in the field of view.

\begin{table*}
         \begin{tabular}{lccccccc}
            \hline
            \noalign{\smallskip}
No.
& R.A.(2000)     
& Dec.(2000)
                 & $S_{0.3-10}$
                 & $S_{0.3-1}$
                 & $S_{1-2}$
                 & $S_{2-10}$
                 & Notes \\
&&&($10^{-4}$ ct s$^{-1}$)& ($10^{-4}$ ct s$^{-1}$) 
& ($10^{-4}$ ct s$^{-1}$) & ($10^{-4}$ ct s$^{-1}$) & \\
            \noalign{\smallskip}
            \hline
            \noalign{\smallskip}
01 & $22~57~12.46$ & $-41~02~32.4$ 
   & $9.2\pm1.4$ & $4.8\pm1.0$ & $3.9\pm0.9$& $<0.5$& \\
02 & $22~57~13.90$ & $-41~03~26.3$ & $2.4\pm0.8$ & $0.6\pm0.3$ 
   & $1.2\pm0.5$ & $0.8\pm0.4$& \\
03 & $22~57~13.94$ & $-41~03~42.6$ & $3.7\pm1.0$ & $2.0\pm0.7$ 
   & $1.2\pm0.5$ & $<0.5$ & \\
04 & $22~57~14.13$ & $-41~02~49.0$ & $6.8\pm1.3$ & $3.6\pm0.9$ 
   & $2.3\pm0.7$ & $0.8\pm0.4$ & \\
05 & $22~57~16.26$ & $-41~02~19.4$ & $3.6\pm0.9$ & $<0.5$ 
   & $0.8\pm0.4$ & $2.7\pm0.8$ & \\
06 & $22~57~17.99$ & $-41~04~11.1$ & $4.9\pm1.1$ & $4.9\pm1.1$ 
   & $<0.5$ & $<0.5$ & SSS\\
07 & $22~57~18.94$ & $-41~06~36.3$ & $6.3\pm1.3$ & $<0.5$ 
   & $3.4\pm0.9$ & $2.3\pm0.8$ & \\
08 & $22~57~19.07$ & $-41~05~00.3$ & $9.1\pm1.4$ & $2.5\pm0.7$ 
   & $3.8\pm0.9$ & $1.7\pm0.6$ & \\
09 & $22~57~22.79$ & $-41~01~51.8$ & $2.4\pm0.7$ & $2.3\pm0.7$ 
   & $<0.5$ & $<0.5$ & SSS\\
10 & $22~57~22.92$ & $-41~04~47.3$ & $2.1\pm0.7$ & $<0.5$ 
   & $1.3\pm0.5$ & $0.8\pm0.4$ & \\
11 & $22~57~23.26$ & $-41~01~15.4$ & $74.8\pm4.0$ & $31.3\pm2.6$ 
   & $34.2\pm2.7$ & $9.4\pm1.4$& BHC \\
12 & $22~57~23.65$ & $-41~01~00.9$ & $3.0\pm0.8$ & $1.4\pm0.6$ 
   & $1.3\pm0.5$ & $<0.5$ & \\
13 & $22~57~23.86$ & $-41~02~55.3$ & $46.3\pm3.2$ & $13.5\pm1.7$ 
   & $21.7\pm2.2$ & $11.3\pm1.6$ & QSO?\\
14 & $22~57~24.62$ & $-41~05~26.3$ & $3.7\pm1.0$ & $<0.5$ 
   & $0.8\pm0.4$  & $1.9\pm0.7$ & \\
15 & $22~57~24.71$ & $-41~03~43.9$ & $318.9\pm8.3$ & $123.0\pm5.2$ 
   & $135.3\pm5.4$ & $56.3\pm3.5$  & ULX\,2\\
16 & $22~57~28.94$ & $-41~02~12.0$ & $493.0\pm10.2$ & $152.9\pm5.7$ 
   & $214.7\pm6.8$ & $122.3\pm 5.1$& ULX\,1\\
17 & $22~57~29.62$ & $-41~02~39.9$ & $17.9\pm2.0$ & $1.4\pm0.6$ 
   & $8.4\pm1.4$ & $7.9\pm1.3$ & QSO\\
18 & $22~57~30.40$ & $-41~02~25.1$ & $17.9\pm2.0$ & $5.1\pm1.1$ 
   & $8.9\pm1.4$ & $3.8\pm0.9$ & \\
19 & $22~57~30.72$ & $-41~02~25.9$ & $8.2\pm1.4$ & $4.7\pm1.0$ 
   & $2.2\pm0.7$ & $1.1\pm0.4$ & SN\,2001ig\\
20 & $22~57~30.76$ & $-41~03~10.8$ & $2.9\pm0.8$ & $0.8\pm0.4$ 
   & $1.2\pm0.5$ & $1.0\pm0.5$ & \\
21 & $22~57~33.06$ & $-41~03~56.6$ & $2.4\pm0.8$ & $<0.5$ 
   & $1.0\pm0.5$ & $1.7\pm0.6$ & \\
22 & $22~57~35.79$ & $-41~05~05.6$ & $6.6\pm1.2$ & $5.1\pm1.1$ 
   & $1.9\pm0.6$ & $<0.5$ & fg star?\\
   \noalign{\smallskip}
            \hline
         \end{tabular}
      \caption{Point-like X-ray sources  
detected at $> 3$-$\sigma$ significance in the $0.3$--$10$ keV band 
inside the D$_{25}$ ellipse of NGC\,7424. The count rate is from 
the two combined observations ($47.32$ ks).}
         \label{t:allsources}
\end{table*}

\begin{table*}   
         \begin{tabular}{lcccccc}
            \hline
            \noalign{\smallskip}
No. & ID & Obs.
                 & $S_{0.3-10}$
                 & $S_{0.3-1}$
                 & $S_{1-2}$
                 & $S_{2-10}$\\
&&&($10^{-4}$ ct s$^{-1}$)& ($10^{-4}$ ct s$^{-1}$) 
& ($10^{-4}$ ct s$^{-1}$) & ($10^{-4}$ ct s$^{-1}$) \\
            \noalign{\smallskip}
            \hline
            \noalign{\smallskip}
11 & BHC   & 1 & $71.2\pm5.5$ & $30.2\pm3.6$ & $30.7\pm3.6$& $10.3\pm2.1$\\
   &       & 2 & $78.7\pm5.7$ & $31.8\pm3.7$ & $37.1\pm3.9$& $8.5\pm1.9$\\
15 & ULX\,2 & 1 & $39.0\pm4.0$ & $15.3\pm2.6$ & $12.3\pm2.3$& $11.1\pm2.1$\\
   &       & 2 & $593.5\pm15.7$ & $228.6\pm9.8$ &$256.1\pm10.3$ 
   &$102.4\pm6.5$ \\
16 & ULX\,1 & 1 & $396.4\pm13.0$ & $139.5\pm7.7$& $170.0\pm8.5$& $81.5\pm5.9$\\
   &       & 2 & $587.8\pm15.7$ & $161.6\pm8.2$& $257.2\pm10.4$ 
   & $164.8\pm8.3$ \\
19 & SN\,2001ig & 1 & $10.7\pm2.1$& $6.4\pm1.7$ &$3.2\pm1.1$ &$1.1\pm0.6$ \\
   &            & 2 & $5.4\pm1.4$& $3.3\pm1.2$& $1.0\pm0.6$& $1.2\pm0.7$\\
   \noalign{\smallskip}
            \hline
         \end{tabular}
      \caption{For selected sources, count rates in the first 
and second {\it Chandra}/ACIS 
observation (separated by about 20 days, see Table 1).}
         \label{t:selsources}
\end{table*}


Assuming a power-law spectrum with photon index 
$\Gamma = 1.7$ and line-of-sight Galactic 
column density $N_{\rm H} = 1.3 \times 10^{20}$ cm$^{-2}$ 
(Dickey \& Lockman 1990), we estimate that the completeness limit 
in the combined image is $\approx 3 \times 10^{37}$ 
erg s$^{-1}$ (i.e., $\approx 2.5 \times 10^{-4}$ 
ct s$^{-1}$).
From the {\it Chandra} Deep Field $\log N$--$\log S$ curve 
(Moretti et al.~2003), we estimate that $\approx 10$ of 
the 22 {\it Chandra} sources (Table 2) are background AGN.
For example, source No.~17 is almost certainly a background 
radio quasar (Ryder et al.~2004): this is also consistent 
with its hard and heavily absorbed power-law spectrum 
in the X-ray band. Source No.~13 is also likely 
to be a background AGN, based on its hard 
X-ray spectrum (not shown here) and its point-like optical 
counterpart with an optical/X-ray flux 
ratio $\sim 1$ (Barger et al.~2003).

The two brightest sources (Table 2) are in the ULX 
regime, and will be discussed in detail later 
(Sections 3.1 and 3.2). The original target 
of the observation, SN\,2001ig, was seen with a count rate 
of $10.7\pm 2.1$ ct s$^{-1}$
in the 2002 May 21--22 observation,
as previously reported (Schlegel \& Ryder 2002). 
This corresponds to a $0.3$--$10$ keV luminosity 
$\approx 6 \times 10^{37}$ 
erg s$^{-1}$ for a thermal plasma model 
with $kT \approx 0.5$ keV (suitable to the soft 
X-ray colors of the source). It declined to 
about half this value in the 2002 June 11 
observation (Table 3).

NGC\,7424 is a bulgeless late-type spiral galaxy 
with a bright, massive nuclear star cluster 
($M_I \approx -11.5$ mag, $M \approx 10^6 M_{\odot}$:  
Walcher et al.~2005). It is still unclear whether 
this subclass of spiral galaxies also have a supermassive 
or perhaps intermediate-mass nuclear BH. 
In some cases, an X-ray source 
coincides with the nuclear star cluster 
(e.g., in M\,33 and in M\,74). Instead, we do not find 
any nuclear X-ray sources in NGC\,7424.
Interestingly, however, a super-soft source  
is located $\approx 290$ pc north-west of the nuclear 
star cluster. For a blackbody temperature 
$kT_{\rm bb} \approx 70$ eV, typical of this class 
of sources, the unabsorbed luminosity is $\approx 8 \times 10^{37}$ 
erg s$^{-1}$ in the $0.3$--$10$ keV band, 
corresponding to a bolometric luminosity 
$\approx 2 \times 10^{38}$ erg s$^{-1}$. 
Another, fainter super-soft source is located 
in the outskirts of the galaxy.

A bright X-ray source (No.~11 in Table 2), located 
in a spiral arm, has an X-ray spectrum 
which is well fitted by a disc-blackbody 
model (Shakura \& Sunyaev 1973; Makishima et al.~1986) 
with $kT_{\rm dbb} = 0.7 \pm 0.1$ keV (Figure 2, 
and Table 4 for the fit parameters). Based on this standard 
model, the unabsorbed luminosity in the $0.3$--$10$ keV band 
is $5^{+5}_{-2} \times 10^{38}$ erg s$^{-1}$.
Spectral properties and luminosity suggest 
that this source is consistent with 
a BH X-ray binary in a high/soft state, 
for example similar to LMC X-1 or LMC X-3 
(Nowak et al.~2001; Wu et al.~2001).

\subsection{ULX\,1: power-law dominated}

In both observations, the spectrum of ULX\,1 
is well fitted (Figure 3) by an absorbed\footnote{Throughout 
this paper, we used {\tt wabs} in {\small {XSPEC}} 
to model the photoelectric absorption 
(Ba\l{}uci\'nska-Church \& McCammon 1992). 
As a check, we repeated our fits with a more complex 
absorption model, {\tt tbabs} (Wilms, Allen \& McCray 2000): 
the difference in the best-fitting value of the column 
density between the two models is $< 10\%$, well within 
the errors, and all the other spectral parameters remain 
essentially unchanged. Besides, the true metal abundance 
of the interstellar medium in NGC\,7424 is unknown.  
Hence, for our current purposes, {\tt wabs} is a suitable 
approximation.} 
power-law of photon index 
$\Gamma \approx 1.8$ (Table 5). The unabsorbed luminosity 
appears to have increased by almost a factor of two 
in the second observation, 
reaching $\approx 9 \times 10^{39}$ erg s$^{-1}$ in 2002 June 
(Figures 4, 5).
There is no evidence of a soft excess or thermal 
disc component. We also coadded the spectra from May and June 
to improve the signal-to-noise ratio of possible thermal-plasma 
emission lines, but found none. A very luminous source 
in the same field was also found by {\it ROSAT}/PSPC 
(observed on 1990 November 11--12), at a count 
rate of $0.036 \pm 0.015$ ct s$^{-1}$ 
({\it ROSAT} All Sky Survey Faint Source Catalog: Voges et al.~2000; 
data available on-line), corresponding 
to an unabsorbed luminosity of $(2\pm1) \times 10^{40}$ 
erg s$^{-1}$ in the $0.3$--$10$ keV band. The displacement between the 
{\it ROSAT} and {\it Chandra} positions is consistent 
with the PSPC error.

Many bright ULXs have been found to have power-law dominated 
spectra, with $1.6 < \Gamma < 2.5$ 
(Winter, Mushotzky \& Reynolds 2006; Stobbart, Roberts \& Wilms 2006), 
similar to this source. In other bright ULXs 
detected with better signal-to-noise, 
a small ``soft excess'' with fitted blackbody temperatures 
$\sim 0.1$--$0.2$ keV (sometimes interpreted as the disc 
of an intermediate-mass BH: Miller, Fabian \& Miller 2004) 
is often also present, in addition to the dominant power-law. 
In the standard understanding of BH spectral states 
(e.g., McClintock \& Remillard 2006), the power-law 
is produced via inverse-Compton scattering 
of cooler seed photons in a hot corona.
However, the hardness of the ULX power-law component  
is difficult to reconcile 
with the standard spectral-state classification 
of Galactic BH candidates. 
The latter class of sources is generally power-law dominated  
either in the low/hard state---but then, 
their X-ray luminosity is only $\la 0.01 L_{\rm Edd}$---or in 
the very high state, at $L_{\rm X} \sim L_{\rm Edd}$---but 
in that case, the power-law 
is steeper, with a photon index $\Gamma \ga 2.5$.

Various scenarios have been suggested 
to explain hard power-law dominated ULXs.
They might indeed be in the low/hard state, 
emitting well below their Eddington limit 
(Winter et al.~2006). 
If that is the case, their BH masses would be 
as high as $\sim$ a few $10^3 M_{\odot}$, 
supporting the intermediate-mass BH interpretation.
However, it is not clear why we have never seen 
any of those objects in a high state, 
with $L_{\rm X} \ga 10^{41}$ erg s$^{-1}$.


Alternatively, it was speculated that the true X-ray spectrum 
in the $0.3$--$10$ keV band is not really a power-law, 
but the sum of a hot disc ($kT_{\rm dbb} \sim 1$--$2$ keV) 
and an additional, cooler ($kT \sim 0.2$ keV) thermal 
component produced via Compton downscattering 
(Stobbart et al.~2006). In many cases, it is hard to distinguish 
between a power-law and a disc-blackbody component 
in the limited {\it Chandra} energy band, particularly 
for spectra with only $\sim 1000$ counts. In the case 
of ULX\,1, in particular, we also obtain formally good 
fits with a thermal component at $kT_{\rm bb} \approx 0.20$ keV 
and a disc-blackbody component with 
$kT_{\rm dbb} = 1.20 \pm 0.08$ keV in the first 
observation, and $kT_{\rm dbb} = 1.44 \pm 0.09$ keV in the second one. 
This alternative model, however, does not 
improve the simpler power-law fit. We do not find evidence 
of spectral curvature at energies $\ga 2$ keV, 
as noted in other sources (Stobbart et al.~2006), 
but the signal-to-noise level is too low to draw 
any firm conclusions.

A third possibility is that such power-law-dominated ULXs
belong to the same steep-power-law state of Galactic BHs 
(assuming that we can indeed apply the same spectral-state 
classification to stellar-mass BHs and ULXs). 
The fitted photon index may appear flatter perhaps because 
of very broad, relativistically-smeared absorption lines 
from highly ionized plasma, which remove some of the 
power-law emission at lower energies (Goncalves \& Soria 2006).

A fourth possibility is that the X-ray power-law component 
is not the result of thermal Comptonization of disc photons 
in a corona, but is produced independently, via other mechanisms.
For example, it could result from nonthermal synchrotron 
and Compton processes in a relativistic jet
(Markoff, Falcke \& Fender 2001; Markoff, Nowak \& Wilms 2005).
It has been proposed (Kuncic \& Bicknell 2004) 
that magnetized disc outflows can drain 
most of the accretion power from the disc, because 
of non-zero magnetic torques acting on its surface. 
If we are observing such sources along 
the jet axis, the thermal disc component becomes 
negligible compared with the boosted power-law component, 
even for accretion rates $\sim$ Eddington rate. 
This would support the microquasar scenario for ULXs.

\begin{center}
   \begin{table}
      \caption[]{Best-fit parameters of an absorbed disc-blackbody 
model, fitted to the {\it Chandra}/ACIS spectrum 
of source No.~11 (a BH candidate). 
The spectrum is coadded from the 2002 May and June data. 
The quoted errors are the 90\% confidence limit and
$N_{\rm H,Gal} = 1.3 \times 10^{20}$ cm$^{-2}$.} 
         \label{table1a}
\begin{center}
         \begin{tabular}{lr}
            \hline
            \hline
            \noalign{\smallskip}
            Parameter    & Value (May/June combined)   \\[2pt]
            \noalign{\smallskip}
            \hline
            \hline
            \noalign{\smallskip}
            \noalign{\smallskip}
                $N_{\rm H}~(10^{20}~{\rm cm}^{-2})$ &
                        $<2.4$\\[2pt]
                $kT_{\rm dbb}~({\rm keV})$  &  $0.69^{+0.09}_{-0.09}$ \\[2pt]
                $K_{\rm dbb}~(10^{-3})$ &
                        $7.2^{+4.8}_{-2.7}$ \\
            \noalign{\smallskip}
            \hline
            \noalign{\smallskip}
                $\chi^2_\nu$ & $0.96\,(18.2/19)$ \\[2pt]
                $f_{0.3{\rm -}10}^{\rm obs}
                         ~(10^{-14}~{\rm erg~cm}^{-2}~{\rm s}^{-1})$
                        &  $3.1^{+3.3}_{-1.3}$ \\[2pt]
                $f_{0.3{\rm -}10}^{\rm em}
                         ~(10^{-14}~{\rm erg~cm}^{-2}~{\rm s}^{-1})$
                        &  $3.3^{+3.3}_{-1.4}$ \\[2pt]
                $L_{0.3{\rm -}10}~(10^{38}~{\rm erg~s}^{-1})$ 
                        & $5.1^{+4.9}_{-2.1}$ \\
            \noalign{\smallskip}
            \hline
         \end{tabular}
\end{center}
   \end{table}
\end{center}

\begin{center}
   \begin{table}
      \caption{Best-fit parameters for a power-law fit 
to the {\it Chandra}/ACIS spectra of ULX\,1 
in the 2002 May and June observations. As before,
the quoted errors are the 90\% confidence limit and
$N_{\rm H,Gal} = 1.3 \times 10^{20}$ cm$^{-2}$.} 
         \label{table1b}
\begin{center}
         \begin{tabular}{lrr}
            \hline
            \hline
            \noalign{\smallskip}
            Parameter    & Value (Obs 1) & Value (Obs 2)  \\[2pt]
            \noalign{\smallskip}
            \hline
            \hline
            \noalign{\smallskip}
            \noalign{\smallskip}
                $N_{\rm H}~(10^{20}~{\rm cm}^{-2})$ &
                        $7.3^{+4.0}_{-3.8}$
                        & $12.3^{+3.6}_{-3.3}$ \\[2pt]
                $\Gamma$  &  $1.90^{+0.18}_{-0.16}$ &
                        $1.73^{+0.10}_{-0.12}$\\[2pt]
                $K_{\rm po}~(10^{-5})$ &
                        $5.4^{+0.9}_{-0.8}$ & $8.4^{+1.1}_{-1.0}$\\
            \noalign{\smallskip}
            \hline
            \noalign{\smallskip}
                $\chi^2_\nu$ & $0.72\,(33.9/47)$ & $0.83\,(59.9/72)$  \\[2pt]
                $f_{0.3{\rm -}10}^{\rm obs}$
                        $~(10^{-13}~{\rm CGS})$
                        &  $2.7^{+0.1}_{-0.3}$ & $4.6^{+0.3}_{-0.3}$\\[2pt]
                $f_{0.3{\rm -}10}^{\rm em}$
                        $ ~(10^{-13}~{\rm CGS})$
                        &  $3.2^{+0.2}_{-0.2}$ & $5.7^{+0.2}_{-0.3}$\\[2pt]
                $L_{0.3{\rm -}10}~(10^{39}~{\rm erg~s}^{-1})$ 
                        & $5.1^{+0.3}_{-0.3}$ & $8.9^{+0.4}_{-0.3}$\\
            \noalign{\smallskip}
            \hline
         \end{tabular}
\end{center}
   \end{table}
\end{center}

\begin{figure}
\epsfig{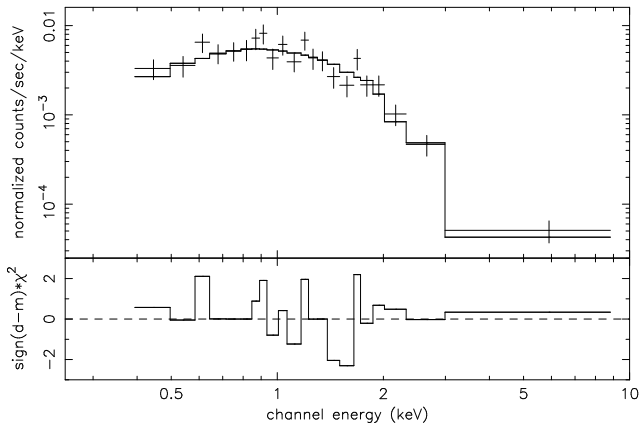}
\caption{{\it Chandra}/ACIS X-ray spectrum 
of a BH candidate (source No.~11) in a high/soft state, fitted 
here with an absorbed {\tt diskbb} 
model, with $kT_{\rm dbb} = 0.7\pm 0.1$ 
keV. See Table 4 for the best-fitting parameters.}
\end{figure}

\begin{figure}
\epsfig{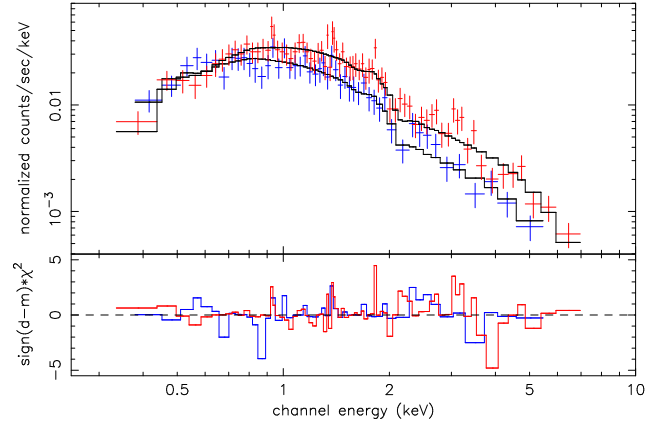}
\caption{{\it Chandra}/ACIS X-ray spectra 
of ULX\,1 in the 2002 May and June observations 
(blue and red datapoints, respectively). 
Both spectra can be well fitted with absorbed power-law models. 
See Table 3 for details of the best-fitting parameters.}
\end{figure}

\begin{figure}
\epsfig{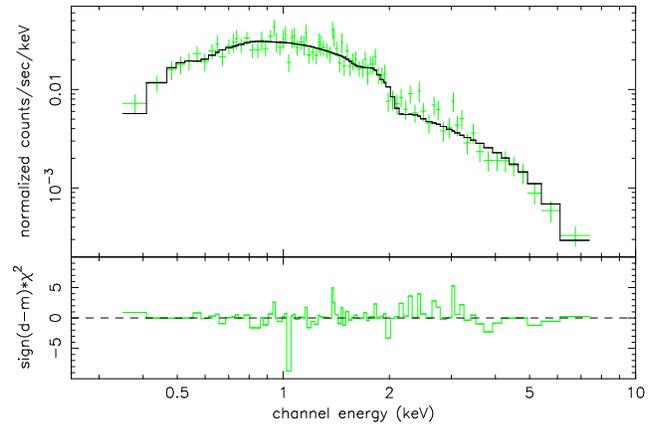}
\caption{Coadded {\it Chandra}/ACIS X-ray spectrum 
of ULX\,1, fitted with an absorbed power-law model (photon index 
$\Gamma = 1.8\pm 0.1$).}
\end{figure}

\begin{figure}
\epsfig{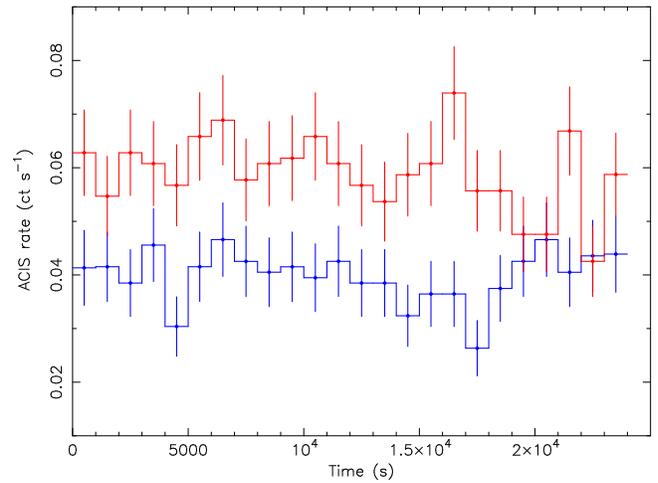}
\caption{{\it Chandra}/ACIS lightcurve ($0.3$--$10$ keV band) 
for ULX\,1, during the two observations. The lower (blue) curve 
is from 2002 May, the upper (red) curve from 2002 June.}
\end{figure}

\subsection{ULX\,2: catching a state transition}

The second ULX in NGC\,7424 was an order of magnitude 
brighter on 2002 June 11, compared 
with three weeks earlier (Table 3; Figures 6, 7). In the first 
observation, the source has a hard spectrum, 
consistent with a power-law of index $\Gamma = 1.4 \pm 0.4$, and  
unabsorbed luminosity $\approx 5 \times 10^{38}$ erg 
s$^{-1}$ (Table 6). 
In the second observation, its spectrum 
shows a softer continuum, with  
photon index $\Gamma = 2.2 \pm 0.2$ 
if fitted with a simple power-law model 
($\chi^2_{\nu} = 46.3/54$).
In fact, there are weak but systematic features 
at energies $\approx 0.7$--$1.3$ keV, typical 
of line emission from an optically-thin thermal plasma;
adding a thermal component improves the fit 
($\chi^2_{\nu} = 31.7/50$). In this model (Table 6), 
$\Gamma = 1.8^{+0.2}_{-0.1}$, 
the plasma temperature is $1.0\pm0.2$ keV, the unabsorbed 
luminosity in the thermal component 
is $\approx 8 \times 10^{38}$ 
erg s$^{-1}$ (mostly in the $0.3$--$2$ keV band), 
and the total unabsorbed luminosity is 
$\approx 6.5 \times 10^{39}$ erg s$^{-1}$.
Comparing the fluxes and spectral models for the two 
observations, we rule out the possibility that
the thermal-plasma component is simply underlying 
emission from unrelated sources, for example 
a supernova remnant (SNR), 
because it would have already been detected 
in the first observation.
It could instead be the result of the outburst, 
which may have produced an outflow of hot gas 
around the source.

State transitions 
are very unusual for bright ULXs, which generally 
do not vary by more than a factor of a few over 
many years; a notable exception 
is a source in NGC\,3628 (Strickland et al.~2001). 
On the other hand, stronger variability or on/off 
transitions have been seen in fainter 
($L_{\rm X} \la$ a few $10^{39}$ erg s$^{-1}$) ULXs, 
for example in the Antennae (Fabbiano et al.~2003).
The short timespan ($\la 20$ days) over which 
the spectral change took place is similar to state transitions 
in stellar-mass BHs, usually attributed 
to thermal-viscous disc instabilities 
(Meyer \& Meyer-Hofmeister 1981).  
Such limit-cycle behaviour has never been clearly 
identified in ULXs.

For the thermal-viscous instability to occur, the outer disc 
has to be cooler than the Hydrogen ionization threshold.
This condition can be used to constrain the mass and size 
of the binary system. For example, it was shown 
(Kalogera et al.~2004) that if the Roche-lobe filling 
donor star has a mass $\ga 7 M_{\odot}$ (consistent with 
the young optical environment around ULX\,2, 
see Section 4), transient behaviour is statistically 
more likely to occur for BH masses $\ga 50 M_{\odot}$.
Conversely, for donor stars with masses $\la 7 M_{\odot}$, 
the limit-cycle instability can occur for any BH mass.
With similar arguments, based on the temperature of 
an X-ray irradiated disc, it can be shown (van Paradijs 1996) 
that ULX\,2 should have a limit-cycle behaviour if 
\begin{eqnarray}
  && \left[2.35 \times 10^{11} P^{2/3}_{\rm d} 
  \left(\frac{M_{\rm BH} + M_2}{M_{\odot}}\right)^{1/3} \right]
  \nonumber\\
  && \ \times \left(0.38 + 0.2 \log \frac{M_{\rm BH}}{M_2}\right)   
   \ga 10^{13} \ {\rm cm},
\end{eqnarray}
where $P_{\rm d}$ is the binary period in days, 
and we used the emitted luminosity of ULX\,2 in its 
higher state. Choosing, for example, 
$M_{\rm BH} \approx 50 M_{\odot}$ and $M_2 \approx 20 M_{\odot}$, 
one obtains $P_{\rm d} \ga 100$; for 
$M_{\rm BH} \approx M_2 \approx 10 M_{\odot}$, $P_{\rm d} \ga 250$.

\begin{center}
   \begin{table}
      \caption[]{Best-fit parameters 
to the {\it Chandra}/ACIS spectra of ULX\,2 
in the 2002 May and June observations. 
For the June observation, we fitted 
the source first with a simple power-law, 
then with a power-law plus 
optically-thin thermal plasma, 
{\tt wabs}$_{\rm Gal}~\times$ {\tt wabs} $\times$ ({\tt po} 
$+$ {\tt vapec}) in {\footnotesize {XSPEC}}, 
As before,
the quoted errors are the 90\% confidence limit and
$N_{\rm H,Gal} = 1.3 \times 10^{20}$ cm$^{-2}$.} 
         \label{table1c}
\begin{center}
         \begin{tabular}{lrrr}
            \hline
            \hline
            \noalign{\smallskip}
            Parameter    & Value 1 & Value 2 & Value 2 \\[2pt]
	    &  po & po & po+vapec \\[2pt]
            \noalign{\smallskip}
            \hline
            \hline
            \noalign{\smallskip}
            \noalign{\smallskip}
                $N_{\rm H}~(10^{20}~{\rm cm}^{-2})$ &
                        $<9.5$ & $12.0^{+3.8}_{-3.6}$
                        & $4.3^{+4.4}_{-1.9}$ \\[2pt]
                $\Gamma$  &  $1.43^{+0.47}_{-0.39}$ & 
	                $2.20^{+0.17}_{-0.15}$&
                        $1.84^{+0.23}_{-0.12}$\\[2pt]
                $K_{\rm po}~(10^{-5})$ & 
                        $0.35^{+0.13}_{-0.09}$ & $9.5^{+1.4}_{-1.2}$
	              & $5.8^{+1.0}_{-1.2}$\\[2pt]
	        $kT_{\rm ap}~({\rm keV})$ & - & - & 
	                $1.00^{+0.20}_{-0.16}$\\[2pt]
	        [Mg/H] & - & - & $6.9^{+7.0}_{-6.9}$\\[2pt]
	        [Si/H]$=$[S/H]$=$[Ar/H] 
                         & - & - & $3.5^{+4.2}_{-3.5}$\\[2pt]
                $K_{\rm ap}~(10^{-5})$ & -  & - & $1.6^{+0.6}_{-0.7}$\\
            \noalign{\smallskip}
            \hline
            \noalign{\smallskip}
                $\chi^2_\nu$ & $1.05$ & $0.86$ & $0.63$  \\
	                     & $(6.35/6)$ & $(46.34/54)$ & $(31.7/50)$ \\[2pt] 
                $f_{0.3{\rm -}10}^{\rm obs}$
                         $~(10^{-13}~{\rm CGS})$
                        &  $0.31^{+0.04}_{-0.08}$ &  $3.4^{+0.2}_{-0.3}$
	                & $3.6^{+0.3}_{-0.5}$\\[2pt]
                $f_{0.3{\rm -}10}^{\rm em}$
                        $ ~(10^{-13}~{\rm CGS})$
                        &  $0.32^{+0.06}_{-0.11}$ & $4.9^{+0.4}_{-0.4}$
	                & $4.1^{+0.4}_{-0.2}$\\[2pt]
                $L_{0.3{\rm -}10}~(10^{39}~{\rm erg~s}^{-1})$ 
                        & $0.50^{+0.10}_{-0.17}$ & $7.8^{+0.6}_{-0.6}$
	                & $6.5^{+0.6}_{-0.3}$\\
            \noalign{\smallskip}
            \hline
         \end{tabular}
\end{center}
   \end{table}
\end{center}

\begin{figure}
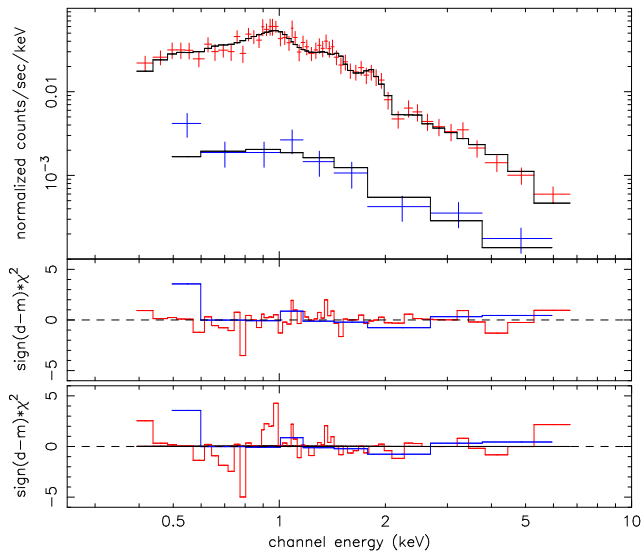

\epsfig{figure=ulx2_apec_fit.ps, angle=270, width=8.3cm}
\epsfig{figure=ulx2_po_fit.ps, angle=270, width=8.4cm}
\caption{{\it Chandra}/ACIS X-ray spectra 
of ULX\,2 in the 2002 May and June observations 
(blue and red datapoints, respectively). 
In its brighter state, ULX\,2 is softer and there is 
a hint of optically-thin thermal-plasma emission 
in addition to a power-law-like component: 
compare the middle panel ($\chi^2$ residuals 
for a power-law plus thermal plasma model), with 
the bottom panel ($\chi^2$ residuals 
for a simple power-law model).
See Table 6 for details of the best-fitting parameters.}
\end{figure}

\begin{figure}
\epsfig{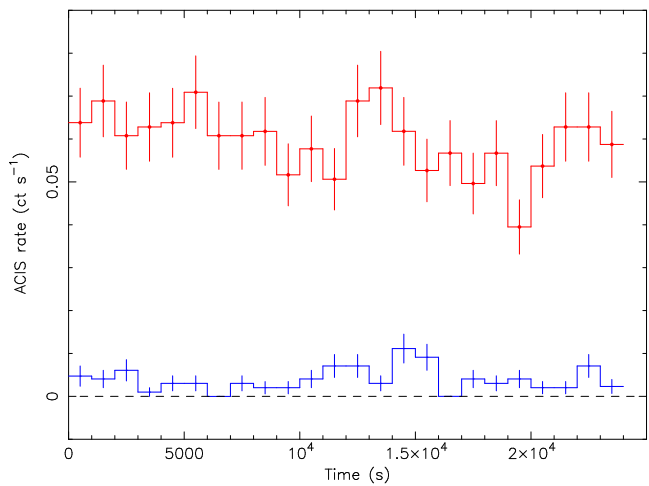}
\caption{{\it Chandra}/ACIS lightcurve ($0.3$--$10$ keV band) 
for ULX\,2, during the two observations. As in Figure 6, 
the blue and red curves are for the first and second observation, 
respectively.}
\end{figure}

\section{Optical environments of the ULXs}

Optical images of the galaxy were obtained 
with the {\it Gemini} Multi-Object Spectrograph 
(GMOS-South), on 2004 September 14, 
in the $r'$ (exposure time $=530$ s), $g'$ ($240$ s) 
and $u'$ ($540$ s) broad bands (Sloan photometric system), 
in exceptional seeing conditions ($\approx 0\farcs35$--$0\farcs45$). 
For more details on the observations, 
as well as a discussion of the extinction, zeropoints 
and colour terms, see Ryder et al.~(2006).
Colour coefficients 
for a simple conversion to the standard $UBVRI$ photometric 
system are given in Table 7 of Smith et al.~(2002).
Ryder et al.~(2006) carried out an optical study 
of the counterpart and environment of SN\,2001ig.
Here, we focus instead on the environments around 
the two ULXs, which appear significantly different.

ULX\,1 is far from massive young star clusters 
or OB associations (Figure 8, top panel). The paucity of bright young stars 
in its environment may be partly explained 
by patchy dust extinction (Ryder et al.~2006); however, 
from the fitted $N_{\rm H}$ in the {\it Chandra} spectrum 
(Table 5) it appears that the X-ray source itself is not affected 
by more than about 1 mag of $V$-band extinction.
There are no optical counterparts formally detected 
above the $3.5\sigma$ level in any of the three {\it Gemini} 
images; however, two faint sources are clearly 
visible inside the error circle in a combined 
image obtained by normalizing and coadding 
the three optical bands (Figure 8, bottom panel). 
For the brighter candidate 
counterpart, we estimate $g' \sim 26$ mag, $r' \sim 26.5$ mag, 
$u' \ga 26$ mag, corresponding to an absolute 
magnitude $-4.5 \la M_V \la -4$ mag, 
typical of main-sequence early B stars.
A few brighter, isolated OB stars (possibly blue 
supergiants) are also found nearby.


Conversely, ULX\,2 is located inside a bright, 
young stellar complex (Figures 9, 10), $\approx 150$ pc in diameter.
This region contains a massive star cluster, 
just south of the positional error circle for ULX\,2, 
and various other fainter groups of OB stars.
The optical brightness of the whole complex 
is $U = (17.99 \pm 0.19)$ mag, $B = (18.81 \pm 0.16)$ mag, 
$V = (18.48 \pm 0.16)$ mag, $R = (18.29 \pm 0.15)$ mag 
(corresponding to $M_V \approx -11.9$ mag). 
For the brightest unresolved cluster, we estimate 
$U = (19.60 \pm 0.19)$ mag, $B = (20.17 \pm 0.16)$ mag, 
$V = (19.77 \pm 0.16)$ mag, $R = (19.54 \pm 0.15)$ mag 
(corresponding to $M_V \approx -10.6$ mag).
For the other smaller groups of stars at the northern  
edge of the X-ray error circle, 
$U = (19.73 \pm 0.19)$ mag, $B = (20.53 \pm 0.16)$ mag, 
$V = (20.48 \pm 0.16)$ mag, $R = (20.45 \pm 0.15)$ mag.
The distinctive average colours of this 
stellar complex, quite blue at $-0.8$ mag $\la U-B \la -0.5$ mag,  
but with $0.1$ mag $\la (B-V) \approx (V-R) \la 0.4$ mag, 
correspond to a brief evolutionary phase 
dominated by stars with masses $\approx 20$--$25$ $M_{\odot}$ becoming 
supergiants. Using the latest evolutionary tracks 
in Starburst99 (Leitherer et al.~1999) 
for instantaneous star formation, 
we estimate an age of $(9.6 \pm 2.0)$ Myr 
at solar metallicity ($Z = 0.02$), or 
$(7.3 \pm 0.9)$ Myr for $Z = 0.008$.
The optical brightness of the whole complex 
corresponds to a total stellar mass 
$4.8 \la \log (M/M_{\odot}) \la 5.2$ at $Z = 0.02$, or
$4.7 \la \log (M/M_{\odot}) \la 4.9$ at $Z = 0.008$. 
For the biggest compact 
cluster, the estimated masses are 
$4.3 \la \log (M/M_{\odot}) \la 4.7$, or 
$4.2 \la \log (M/M_{\odot}) \la 4.4$, 
respectively. The other clusters in the 
stellar complex have masses $\sim 10^3 M_{\odot}$.

It is still not clear whether those two different 
environments for ULX\,1 and 2 
correspond to two fundamentally different classes 
of ULXs, for example in terms of BH mass, age, donor type, 
emission processes, or simply to different 
channels of formation. It has been suggested 
(Portegies Zwart \& McMillan 2002; G\"urkan, Freitag \& Rasio 2004)
that ULXs may contain intermediate-mass BHs formed 
in the core of a young super star-cluster, 
via runaway stellar coalescence; these BHs would  
become X-ray bright if powered for example by Roche-lobe 
accretion from an OB donor star. Notice, however, that 
the young star clusters near ULX\,2 are smaller than 
a super star-cluster, generally defined as having a mass 
$\ga$ a few $10^5 M_{\odot}$. The difference is not purely 
semantic. A super star-cluster is massive and compact enough 
that it may undergo runaway core collapse and coalescence  
of the O stars in its core, over their main-sequence 
life-time, $\sim 3$ Myr. 
Instead, clusters with masses of $\sim 10^3$--$10^4 M_{\odot}$ 
do not have enough O stars to produce an intermediate-mass 
BH through this process; however, they might still be able to form 
a massive progenitor star in the core, via large-scale gas 
collapse and {\it proto}stellar mergers, over a time-scale 
of $\la 2 \times 10^5$ yr. (A similar process may be 
currently happening in our own backyard, in the collapsing 
protocluster NGC\,2264-C; Peretto, Andr\'e \& Belloche 2006).
In fact, many more ULXs have been found associated to medium-size 
clusters and OB associations, than inside or close 
to super star-clusters. This may already give us 
a clue to their most likely formation process.


\section{Radio observations}

\subsection{ULX counterparts: main results}

Ryder et al.~(2004) monitored SN\,2001ig from 2001 December to 2004 April 
with the ATCA in Narrabri, New South Wales.
The observed field includes the locations of both ULX\,1 and ULX\,2.
A total of 37 observations were taken, with a typical synthesis time of
2--4\,h.
Calibrated visibilities of each of the  observations  were used to 
create sensitive radio images at 1.384, 2.368/2.496, 4.790 and 8.640 GHz.
The {\small{CLEAN}}-ing procedure was restricted to an area containing the
supernova and nearby radio source and was stopped either after 1000 iterations
or when the peak residual reached a level of 0.5\,mJy.
Self-calibration was applied to enhance the dynamic range of the images, 
which were then primary-beam corrected.
We coadded the separate images from each observation at each observing
frequency using the {\small{MIRIAD}} task {\tt{imcomb}}, with 
inverse variance weighting; we chose not to include 
a handful of images that had very sparse (U, V) coverage and
consequently a very elongated beam.

The 2.368/2.496\,GHz data cube was collapsed into a single averaged plane
using the {\small{MIRIAD}} task {\tt{avmaths}}: 
the average frequency was 2.432\,GHz.
However, we obtained a considerable improvement in the signal-to-noise 
ratio of the 2.4 GHz emission by omitting the 2.496-GHz data taken 
until 2002 June 30, because they had higher in-band interference.
The root-mean-square noise level in the four bands is listed
in Table~\ref{t:radio}. This is the deepest multi-band search 
of ULX radio counterparts in a nearby galaxy, 
with excellent sensitivity (detection limit $\sim 0.1$ mJy) 
especially at the three highest observing
frequencies.

We find no evidence of radio emission near the location of ULX\,1.
At the location of ULX\,2, on the other hand, there is unresolved 
radio emission significantly above the noise level 
at 2.4 and 4.8 GHz (Table 7); at precisely the same 
location, there is also enhanced emission, above 
the average background level, at 1.4 and 8.6 GHz, 
but not strong enough to be considered a formal detection 
($\approx 2.2$--$2.3\sigma$ above the background).
The 2.4 and 4.8 GHz fluxes suggest an optically-thin 
synchrotron source of index\footnote{Here we 
define $S_\nu \propto \nu^\alpha$.} $\alpha \approx -0.7$; 
this spectral shape is also consistent 
with the flux measured at 1.4 and 8.6 GHz.

The source is detected at similar levels both in the 
combined 2.496-GHz dataset between 2001 December and 2002 June, 
and (with better signal-to-noise ratio) in the subsequent 2.368-GHz 
observations. This suggests that it did not arise 
as a result of (or in association with) the X-ray state
transition of 2002 May/June, although we do not have a simultaneous 
X-ray/radio observation at that epoch.
Figure~9 shows the resulting 2.368 GHz contours overlaid on 
a \textit{Gemini} $R$-band image, with \textit{Chandra} error 
circles indicating the locations of X-ray point sources 
in the NE quadrant of the galaxy.
The radio emission near ULX\,2 is the third brightest source 
in this field, after SN\,2001ig and the nearby background
source (most likely a quasar).



\begin{center}
\begin{table}
\begin{center}
\begin{tabular}{c|c|c|c}
\hline \hline
$\nu$ & rms  & $S_\nu$  & $S_\nu$ \\
 (GHz) & ($\mu$Jy) & ($\mu$Jy) & ($\mu$Jy)\\
\hline
 &  & Source near ULX\,2 & Source B \\ 
\hline
1.384 & 145 & ($\approx 340$) & ($\la 300$)\\
2.368 & 40 & 235 & 267\\
4.790 & 34 & 138 & 223\\
8.640 & 45 &  ($\approx 100$) & NA\\
\hline\hline
\end{tabular}
\end{center}
\caption{Observing frequency, noise limit and radio flux density measurements
for two bright, unresolved sources: one almost coincident with ULX\,2 
and the other in the southern spiral arm (see text for details).}
\label{t:radio}
\end{table}
\end{center}

Taken together, the 2.4 and 4.8\,GHz ATCA observations of ULX\,2
provide conclusive evidence of a radio counterpart, physically
associated with the X-ray source, the young stellar complex, or both.
There are very few confirmed detections 
of individual radio sources associated with ULXs. 
The full list to-date consists of: one source in NGC\,5408 
(Kaaret et al.~2003; Soria et al.~2006); one in 
Holmberg~II (Tongue \& Westpfahl 1995; Miller, 
Mushotzky \& Neff 2005); one in NGC\,6946 
(Roberts \& Colbert 2006; Swartz et al. 2006 in preparation); 
and probably two in M\,82 (Kronberg \& Sramek 1985; K\"ording 
et al.~2005), where, however, confusion is a problem 
and chance coincidences are more likely.
In NGC\,7424, the centroid of the radio emission, 
both at 2.4 and 4.8 GHz, is clearly offset from the X-ray 
position of ULX 2 by $\approx 1\farcs4$, to the South (Figure~10).
This corresponds to $\approx 80$ pc at the assumed distance.
The offsets cannot be attributed to a systematic astrometric error as
the  radio coordinates of the peak emission from SN\,2001ig and the 
nearby background source lie within their respective \textit{Chandra} 
error circles. On the other hand, considering that there are no other 
radio or X-ray sources within 3 kpc, the possibility of a chance
coincidence at least with the stellar complex, 
if not directly with the ULX, is unlikely.

\subsection{Interpretation}

There are a few possible interpretations of the candidate 
radio counterpart to ULX\,2.
One possibility is that the emission is Doppler-boosted synchrotron
radiation from a jet.
The one-sidedness of the radio offset from the \textit{Chandra} position
is then naturally explained by de-boosting of the opposite jet, as is 
often the case in radio galaxies and quasars.
A radio offset of $\approx 80$ pc is not substantially larger
than the size of the jet in, for example,
the Galactic microquasar SS\,433 (Downes, Pauls \& 
Salter 1986; Dubner et al.~1998). 
However, a flat radio spectrum is expected from 
a steady jet in a beamed source (core emission), whereas our
2.4 and 4.8\,GHz measured fluxes (Table~\ref{t:radio}) imply 
$\alpha \approx -0.7$, which is more consistent with a radio lobe signature.
We can also rule out (see also Freeland et al.~2006) 
the possibility that radio and X-ray emission are produced 
by the same synchrotron component.
Optically-thin, Doppler-boosted radio emission may however 
be produced by sporadic ejections (eg., Fender, 
Belloni \& Gallo 2004). Monitoring of the position and flux 
of the radio source in the next few years would test this scenario.

An alternative possibility is that the emission arises 
in a radio lobe, produced by long-term jet activity from ULX\,2.
It has been suggested 
(Heinz 2002; Hardcastle \& Worrall 2000; Hardcastle 2005) 
that the radio lobes of microquasars may
be considerably less prominent and more short-lived 
than those of FR\,II radio galaxies because
of lower effective densities in the surrounding interstellar medium, 
relative to the intergalactic medium: the under-dense environment 
favours adiabatic losses over synchrotron losses.
Moreover, the counterpart to ULX\,2 appears to be a single radio structure,
whereas FR\,II radio galaxies tend to exhibit double
radio lobes. Nevertheless, it is possible that we see only a single
radio lobe associated with ULX\,2 because only one side of its jet 
has impacted onto a molecular cloud or an over-density in the
local interstellar medium, allowing the lobe to form.

A third possibility is that the radio emission is optically-thin synchrotron
radiation produced by a young SNR, 
perhaps from the progenitor of the BH in ULX\,2 itself.
We already noted from the \textit{Gemini} data (Section 4) that ULX\,2 lies
in a region characterized by young star clusters, 
with characteristic ages $\approx 7$--$10$ Myr.
In fact, the position of the most massive cluster 
is still marginally consistent with the radio source. 
From its estimated mass and age (Section 4), we estimate 
a current rate of 1 SN every $\approx 50$--$100$ kyr 
in that cluster alone, and 1 every $\approx 15$--$30$ kyr 
in the whole complex. Hence, the odds of finding 
a young SNR (age $\la 10^3$ yr) in that region 
are not negligible. As a comparison, the source 
is $\approx 2$ times as luminous as the prototypical 
core-collapse young SNR, Cas A, with a similar spectral index 
(Reynoso \& Goss 2002).
If the SNR is associated with the ULX, 
the radio/X-ray offset can be explained by an initial kick 
velocity: a recoil speed of $\sim 40\,{\rm km \, s^{-1}}$ would 
be sufficient to explain an $\sim 80\,$pc offset after 
$\sim 2 \times 10^6$ yr.

We can estimate an upper limit on the energy content  
of an expanding synchrotron bubble (SNR or lobe) 
using the minimum-energy argument (e.g., Pacholczyk 1970; 
Bicknell 2005).
Since the source in NGC\,7424 is unresolved, its linear 
size must be $\la 2\arcsec$.
Assuming a ratio of relativistic proton to electron energies $\ltapprox 100$,
and a range of electron Lorentz factors 
$1 \ltapprox \gamma_{\rm e} \ltapprox 10^4$,
we obtain a total energy $E_{\rm min} \ltapprox 10^{52}$ erg 
(see also a more extended discussion 
of this argument in Soria et al.~2006).
This exceeds the typical energies inferred for normal 
supernovae: perhaps not
surprisingly, since we already know that its radio luminosity 
would put it at the upper end of all known SNRs.
However, the energy estimate is only an upper limit and 
is most sensitive to the source size $l$, with 
$E_{\rm min} \propto l^{(3\alpha +7)/(\alpha +3)} \simeq l^{2.5}$.
If one could firmly establish that the energy is too high
to have been imparted by a supernova, then the jet scenario 
would become more likely. But one could equally well invoke 
a class of more energetic hypernovae, 
with a higher core mass, as the progenitors of ULXs.
Higher spatial resolution maps  with the VLA (in A or B configuration) could
provide stronger observational constraints on the SNR versus 
lobe interpretation of the radio emission.



\begin{figure}
\epsfig{figure=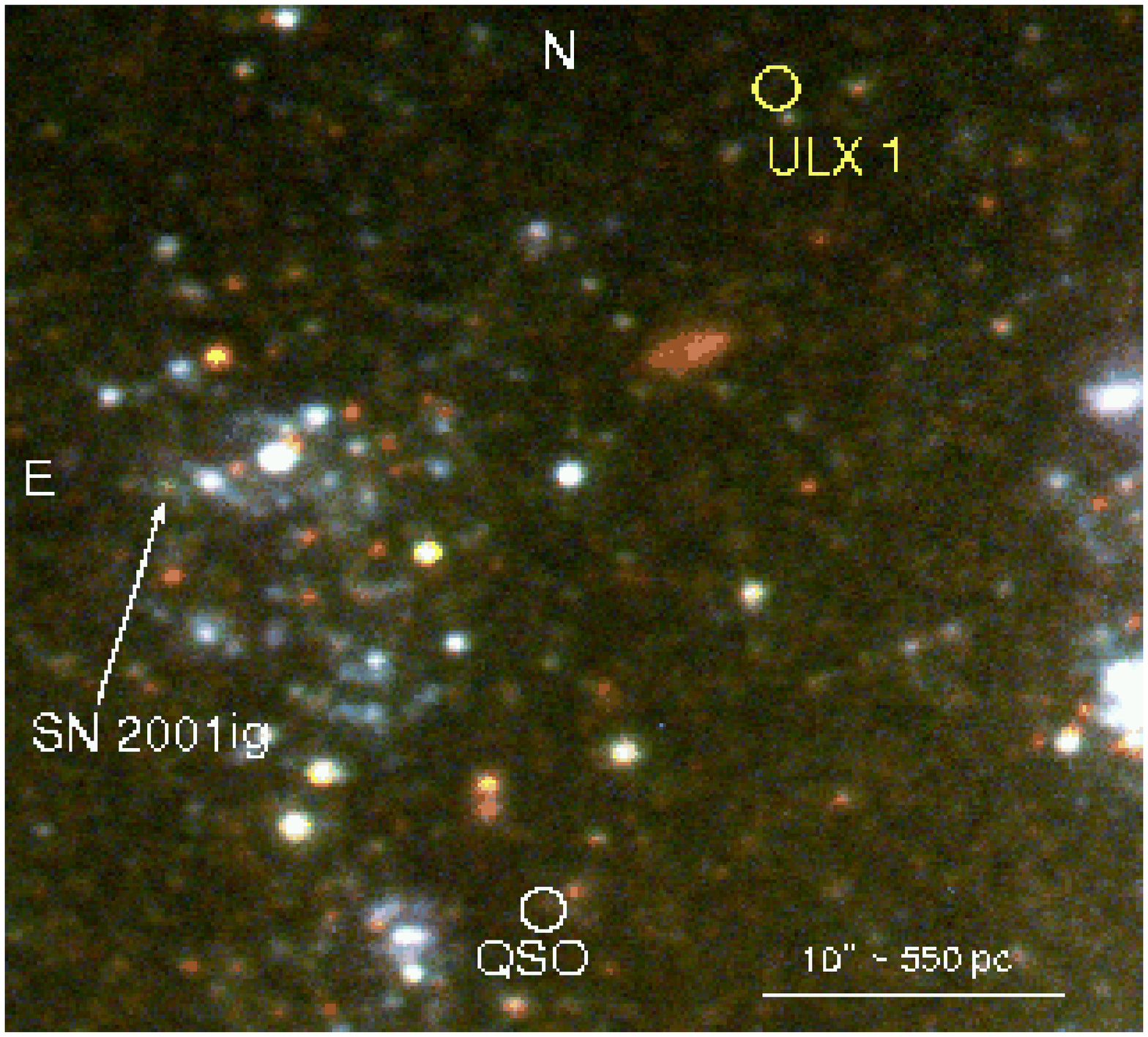, width=8.6cm}
\epsfig{figure=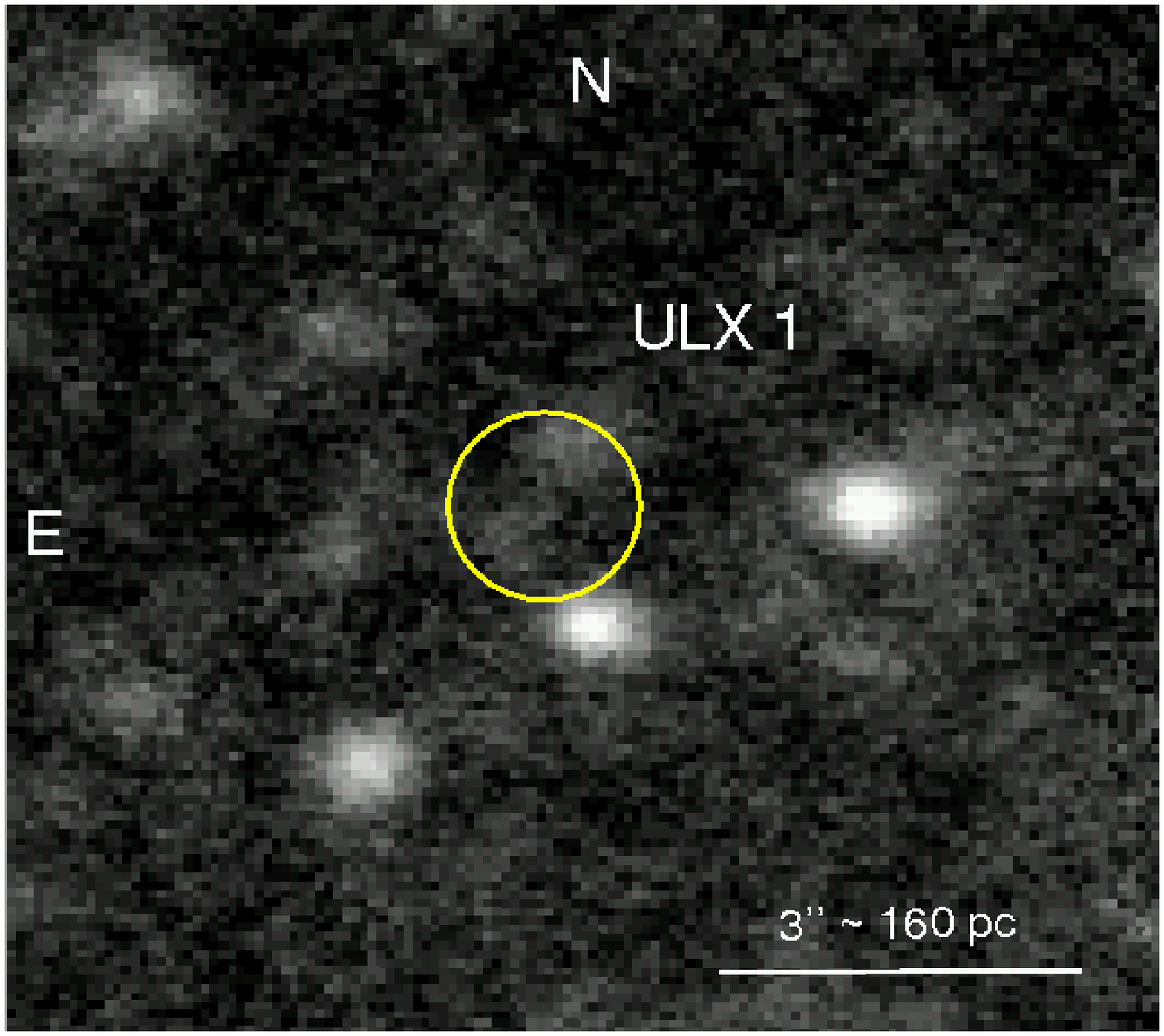, width=8.6cm}
\caption{Top panel: optical environment of ULX\,1, at the outskirts 
of NGC\,7424; true-colour image from our {\it Gemini} observations 
in the $u'$, $g'$ and $r'$ filters. The region around 
SN\,2001ig is dominated by blue and red supergiants, 
but ULX 1 is relatively isolated. Bottom panel: zoomed-in view 
of the region around ULX\,1, in a combined optical image (all three
filters, suitably renormalized and co-added). The two faint stars 
inside the X-ray error circle have an absolute 
brightness $-4.5 \la M_V \la -4$ mag, consistent with 
main-sequence B stars.}
\end{figure}

\begin{figure}
\epsfig{figure=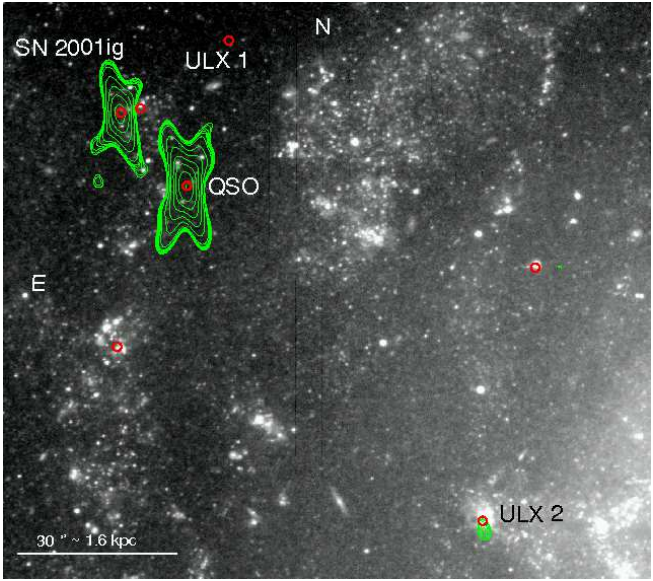, width=8.6cm}
\caption{$2.37$-GHz ATCA radio contours (green) overplotted over a {\it Gemini} 
$r'$-band image. The positions of the {\it Chandra} X-ray sources 
are also marked (red circles).
As it is apparent for the two brightest sources, 
the ATCA beam has a cross shape 
because it is a combination of data taken at various array configurations, 
each with their own elongated beam shape and incomplete (U,\,V) coverage, 
and different position angles. However, the cross pattern 
affects only the lowest contours, which are below the noise 
level for the ULX counterpart; the core is approximately 
Gaussian. The full width half maximum of the $2.37$-GHz beam, in 
a Gaussian approximation, is $2\farcs7 \times 4\farcs9$.}
\end{figure}

A similar ambiguity between the SNR and radio lobe scenarios 
was also discussed for the ULX/radio associations in 
Holmberg II (where the lobe interpretation is deemed much 
more likely: Miller et al.~2005) and NGC\,5408 (where 
the lobe scenario was considered slightly more likely: 
Soria et al.~2006). Both radio counterparts are $\approx 3$ times 
less luminous than the source near ULX\,2 in NGC\,7424. 
The radio source near the ULX in NGC\,6946 was initially classified 
as an SNR (MF\,16: van Dyk et al.~1994; Blair, 
Fesen \& Schlegel 2001) 
but is now also being re-evaluated as a possible 
radio-lobe structure, based on its associated  
optical morphology (Swartz et al.~2006, in preparation); 
its radio luminosity and spectral index are almost exactly 
the same as the source in NGC\,7424.
For M\,82 (K\"ording et al.~2005), both 
ULX-D and ULX-F have persistent, compact radio sources 
slightly displaced ($0\farcs8$ and $0\farcs5$ respectively) 
from the {\it Chandra} positions; both radio sources have  
a luminosity a factor of $2$ higher than the radio 
source near ULX\,2 in NGC\,7424. They have been classified 
as SNRs by analogy with other compact sources in M\,82, 
but there are no elements to rule out a lobe origin. 
There was also a transient source near ULX-D 
($0\farcs6$ displacement), reaching a luminosity 
$8$ times higher than the NGC\,7424 source, in 1981, 
but declining and turning off after a few months; 
it can be interpreted either as a beamed flare from 
the ULX itself, or as a new supernova missed 
in the optical band because of extinction.

Perhaps the most striking similarity of all detections 
of candidate ULX radio counterparts 
is the association in all cases with young or actively star-forming 
regions. This suggests that the exceptionally strong radio emission 
is not intrinsic to the ULXs or due to chance geometrical factors 
(Doppler-boosted ejections) but is more likely to have been enhanced 
by the dense environment (radio lobe scenario) or to be 
due to young SNRs. So far, there have been no core radio detections 
of ULX counterparts. Hence, it is also evident that we cannot 
directly compare X-ray and radio luminosities via fundamental-plane 
relations (e.g., Merloni, Heinz \& di Matteo 2003); we may instead  
compare, in principle, the ULX (instantaneous) X-ray luminosities with 
the (integrated) mechanical power injected into the radio lobes.
If some or most ULX radio counterparts are synchrotron lobes, 
we would expect to find similar structures around other ULXs 
currently in quiescence; the relative abundance of such sources 
would constrain the average ULX duty cycle. It would be even 
more difficult to recognize and distinguish those radio 
sources from bright SNRs, when they are not associated 
with a bright, accreting X-ray source. If we only 
had the first of our two {\it Chandra} observations, 
ULX\,2 itself may have gone unnoticed and its radio 
counterpart would have been classified as an SNR without 
further questioning. 

\subsection{Source B: a pulsar wind nebula?}

Finally, we found another exceptionally bright, 
unresolved radio source located $\sim 1\arcmin$ south-west 
of the nucleus, coincident with another bright knot 
of young stars and clusters inside the southern spiral arm 
(Figures 1, 11); in the optical bands, this young complex is  
$\sim 1$ mag brighter than the stellar complex associated with ULX\,2.
A study of that region is beyond the scope of this paper. Here 
we simply point out that there is no X-ray emission 
$> 3 \times 10^{37}$ erg s$^{-1}$ from that complex. Instead, 
the radio source (J2000 coordinates: R.A. $= 22^h 57^m 16^s.19$; 
Dec. $= -41^{\circ} 05\arcmin 17\farcs7$), labelled 
``Source B'' in Table 7, is even brighter than the one near ULX\,2. 
It has a rather flat spectral index ($\alpha \approx -0.25$ 
between $2.37$ and $4.79$ GHz.) Such indices 
are generally found in plerions, 
also known as pulsar-wind-nebulae (Weiler \& Panagia 1978; 
Gaensler \& Slane 2006); 
the Crab Nebula is the best-studied example. The radio emission 
in this class of SNRs is mostly powered by 
spin-down luminosity of the central compact object 
(a pulsar in all known cases), via a magnetized wind 
which may also interact with the expanding SNR shell.
The radio luminosity of Source B in NGC\,7424
is $\sim 10$ times the Crab luminosity; among 
the highest known, although similar or even higher 
luminosities were found in a few recent SNe 
such as SN\,1979C and SN\,1986J 
(Bartel \& Bietenholz 2005). 
If Source B is indeed a pulsar wind nebula, a radio luminosity 
$\approx 3.6 \times 10^{25}$ erg s$^{-1}$ Hz$^{-1}$ 
at $4.79$ GHz
suggests an age $\la$ a few $100$ yr (e.g., 
Bandiera, Pacini \& Salvati 1984; Frail \& Scharringhausen 1997); 
its X-ray luminosity may be $\sim 10^{37}$ erg s$^{-1}$ 
(e.g., Becker \& Tr\"umper 1997), just 
below our detection limit.


\begin{figure*}
\epsfig{figure=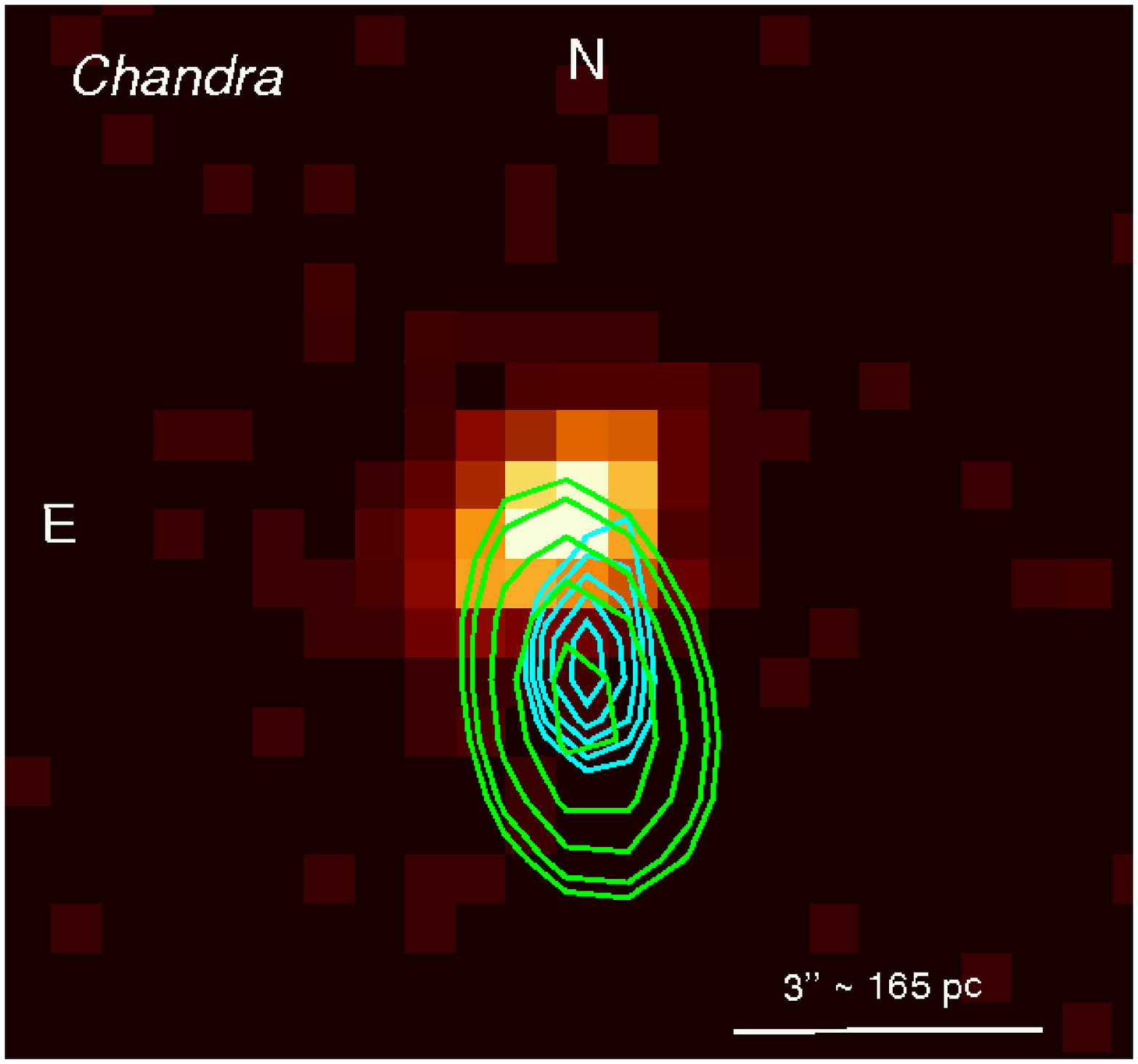, width=4.3cm}
\epsfig{figure=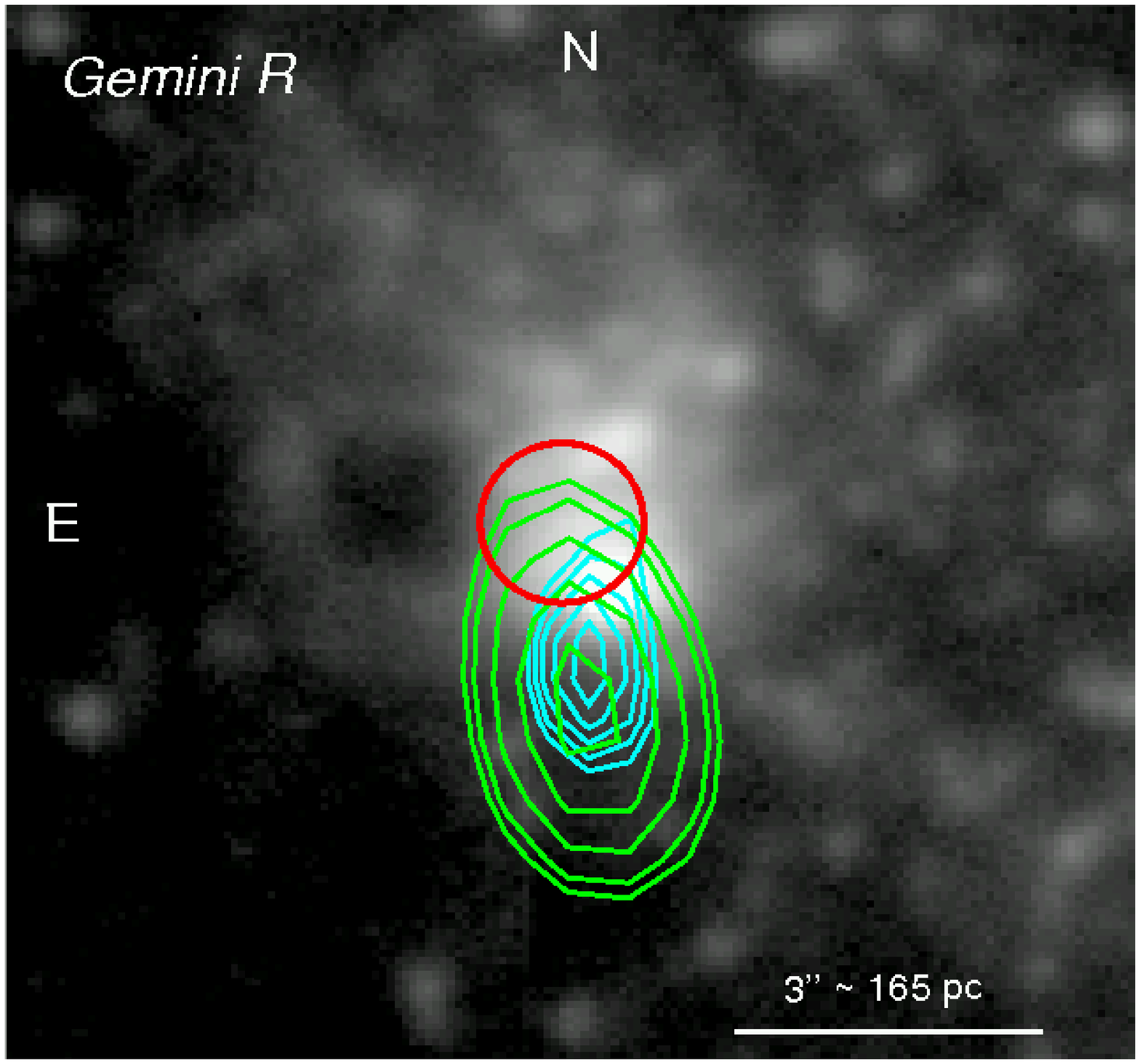, width=4.3cm}
\epsfig{figure=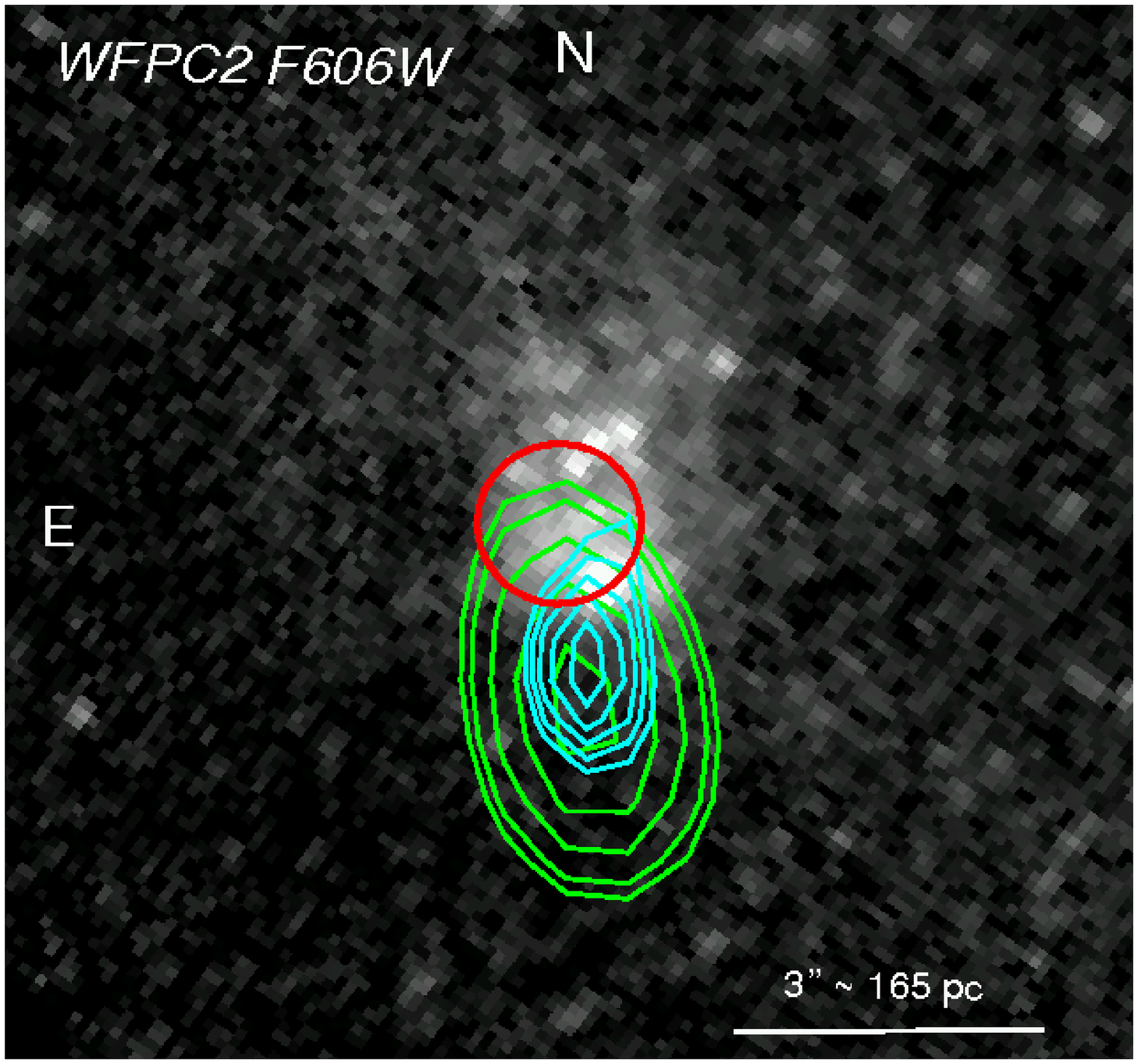, width=4.3cm}
\epsfig{figure=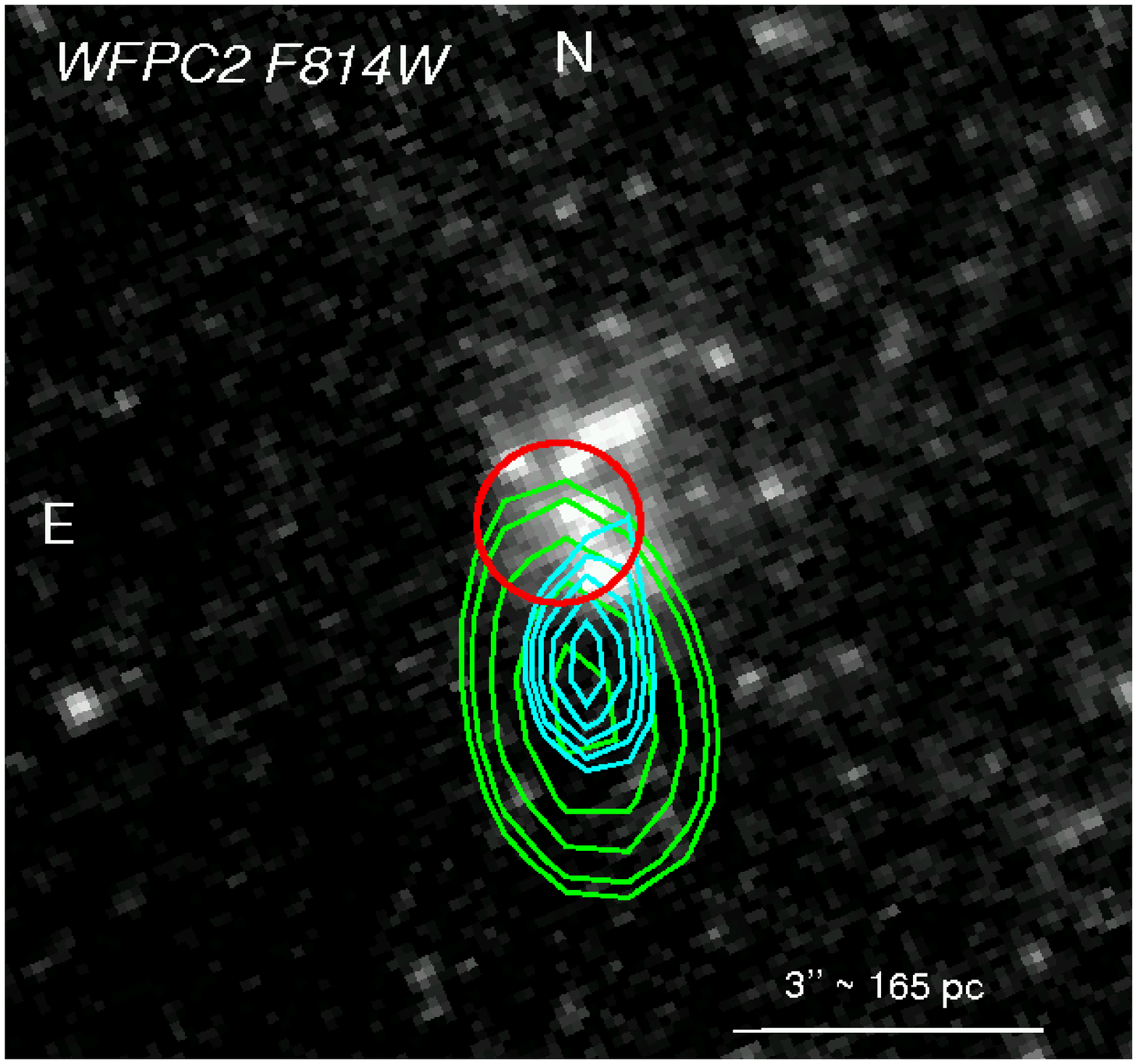, width=4.3cm}
\caption{X-ray, radio and optical associations near ULX\,2. 
The ATCA contours at $2.37$ GHz (green) and $4.79$ GHz (cyan) 
have been superimposed, from left to right, over: 
the {\it Chandra}/ACIS $0.3$--$8$ keV image; 
the {\it Gemini} $r'$-band image; two archival 
snapshot {\it HST}/WFPC2 F606W and F814W images.
The Gaussian full widths half maxima of the radio beams 
are $2\farcs7 \times 4\farcs9$ at $2.37$ GHz 
and $1\farcs4 \times 2\farcs5$ at $4.79$ GHz.
}
\end{figure*}

\begin{figure}
\epsfig{figure=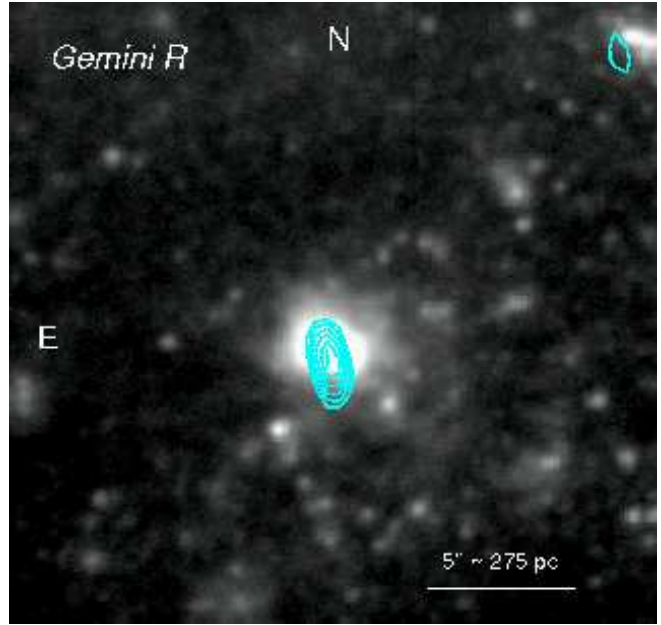, width=8.6cm}
\caption{$4.79$-GHz ATCA radio contours (cyan) overplotted over a {\it Gemini} 
$r'$-band image of another bright young stellar complex, 
$\sim 1$ kpc Southwest of the nucleus. The full width 
half maximum of the beam is $1\farcs4 \times 2\farcs5$.
This radio source is even brighter 
than the one associated with ULX\,2. Based on its flatter 
spectral index, we argue that it is likely to be 
a young pulsar wind nebula. }
\end{figure}

\section{Summary and conclusions}

We studied the brightest X-ray sources and their optical 
and radio counterparts and environments 
in the face-on spiral galaxy NGC\,7424.  
The galaxy was originally monitored in various energy bands 
to study the evolution of SN\,2001ig. In particular, 
we studied it with {\it Gemini} in the $u'$, $g'$ 
and $r'$ optical bands, in 2004; with the ATCA 
in four radio bands, over many observations 
between 2001 and 2004; and 
with {\it Chandra}, twice, on 2002 May 21--22 and June 11. 
Optical and radio results on  
SN\,2001ig are reported elsewhere 
(Ryder et al.~2004, 2006); here we only revise 
its X-ray colours and luminosity, $\approx 6 \times 10^{37}$ 
erg s$^{-1}$ on 2002 May 21--22, declining 
to $\approx 3 \times 10^{37}$ erg s$^{-1}$ on 2002 June 11, 
for a $0.5$ keV thermal plasma.

Our monitoring of NGC\,7424 has led us 
to the serendipitous discovery of two ULXs 
with interesting multi-band properties.
ULX\,1 shows a $75\%$ increase in flux between the two 
observations, reaching an emitted luminosity 
$\approx 9 \times 10^{39}$ erg s$^{-1}$ in the second one. 
It may have been even brighter when observed 
by {\it ROSAT}, in 1990.
Its X-ray spectrum is well fitted with 
a power-law of photon index $\Gamma = 1.8\pm0.1$. This is 
significantly flatter (harder) 
than the typical values found in Galactic BHs when they are bright 
($L_{\rm X} \sim L_{\rm Edd}$) {\it and} power-law dominated. Various 
other bright ULXs have shown this feature, which is yet 
to be properly understood: 
we have briefly discussed some possible explanations. 
ULX\,1 is in a relatively empty 
interarm region, far from any bright clusters or star-forming 
complexes; the brightest optical source in the {\it Chandra} 
error box has an absolute magnitude $M_V \sim -4.5$ mag, 
consistent with a main-sequence B star.

ULX\,2 is more peculiar for at least three reasons. Firstly, 
it showed some kind of outburst or state transition between 
the two {\it Chandra} observations, increasing its brightness 
by an order of magnitude. Thermal plasma emission 
is also detected in its high state. Secondly, unlike ULX\,1, 
it is located in an exceptionally 
bright, young stellar complex (age $\approx 7$--$10$ Myr 
depending on the assumed metal abundance), which also includes 
a few compact clusters. 
It is still unclear why some ULXs are in low-density, 
probably older environments 
while others are associated with OB associations and young star
clusters. A similar dichotomy is sometimes found within 
the same galaxy. For example, the late-type spiral galaxy 
NGC\,4559 also contains two bright ULXs (Cropper et al.~2004), 
one of them located in an interarm region relatively 
devoid of bright stars, the other in a bright 
star-forming complex a few hundred pc in size 
(Soria et al.~2005).
 
In the case of ULX\,2 in NGC\,7424, the biggest star  
cluster in its environment has a mass $\sim$ a few $10^4 M_{\odot}$. 
In general, ULXs seem to be associated 
with young medium-size clusters much more often than 
with super star-clusters ($M \sim 10^6 M_{\odot}$). 
This may give us a clue to the most likely 
formation process for the accreting BHs. 
In fact, the observed X-ray luminosities only require 
BH masses up to $\sim 150 M_{\odot}$. The necessary 
(but not sufficient) condition to end up with such a BH 
is to have a massive stellar progenitor, perhaps 
up to $\sim 300 M_{\odot}$, preferebly with low metal abundance.
We speculate that the most efficient process to create such a massive  
progenitor could be an externally-triggered, dynamical gas collapse 
in a medium-size protocluster, together with a few Class-0 protostellar 
mergers, rather than runaway mergers of main-sequence O stars 
in a super star-cluster.

The third reason why ULX\,2 is remarkable is the presence 
of an exceptionally bright (twice as luminous as Cas A), 
unresolved radio source slightly offset from the ULX 
($\approx 80$ pc). 
Its radio spectral index ($\alpha \approx -0.7$)
is consistent with optically-thin synchrotron emission.
Two alternative interpretations for the radio source 
are either lobe emission, powered by a ULX jet, 
or a young, bright SNR (or perhaps hypernova 
remnant) associated with the young stellar complex 
or maybe even the progenitor of the ULX itself.
Compact radio sources of similar brightness 
and spectral index have been found associated with 
a handful of other ULXs, in some cases implying 
energies up to $\sim 10^{52}$ erg. In almost all cases, 
it is impossible to rule out either of those 
two scenarios altogether. It may even be possible 
that jet lobes and an underlying SNR coexist, like in 
the Galactic microquasar SS\,433.

Another bright (in fact, even brighter), unresolved 
radio source was found in the southern spiral arm, also 
associated with a young, massive stellar complex.
In that case, however, there are no X-ray sources nearby.
Moreover, its radio spectral index is flatter 
($\alpha \approx -0.25$). We speculate 
that it is most likely to be a very young pulsar-wind 
nebula, $\sim 10$ times more luminous than the Crab Nebula. 
However, further radio and X-ray observations of this peculiar 
source will be necessary to test this hypothesis and rule out 
other scenarios, for example beamed radio core emission 
from a microblazar.

This year, {\it Chandra} is carrying out a complete snapshot survey 
of ULXs in about 150 galaxies within 15 Mpc, to pinpoint 
their locations (PI: D. Swartz). A systematic optical survey 
of the same galaxies is also being carried out or planned, 
to classify and study the ULX optical counterparts. 
However, much more work remains to be done to plan and 
implement a correspondingly deep, complete radio survey, 
which would determine what fraction of ULXs are associated 
with optically-thin synchrotron sources, whether 
those sources are statistically different from 
normal SNRs, and perhaps also help distinguishing 
between SNR and radio lobe scenarios. The long-term 
goal of combined radio/X-ray studies is to determine 
the balance of power in accreting BHs, between radiative 
(X-rays) and mechanical luminosity (jet power) 
in different spectral states.


\section*{Acknowledgments}
We thank Geoff Bicknell and Richard Hunstead 
for discussions. RS acknowledges support from an OIF 
Marie Curie Fellowship, as well as financial support 
from the University of Sydney during his visit there.

\bsp


\end{document}